\documentclass[a4paper, amsfonts, amssymb, amsmath, reprint, showkeys, nofootinbib, twoside, aps, pra]{revtex4-1}

\usepackage[english]{babel}
\usepackage[utf8]{inputenc}
\usepackage[T1]{fontenc}
\usepackage{lipsum}
\usepackage{csquotes} 

\usepackage{amsmath, amssymb, amsthm}
\usepackage{algorithm}
\usepackage[noend]{algpseudocode}
\usepackage{mathtools}
\usepackage{physics}
\usepackage{blkarray}
\usepackage{siunitx}
\usepackage[version=4]{mhchem} 
\usepackage{qcircuit} 

\usepackage{xcolor}
\usepackage{graphicx} 
\usepackage[left=23mm,right=13mm,top=35mm,columnsep=15pt]{geometry}
\usepackage{adjustbox}
\usepackage{placeins}
\usepackage[normalem]{ulem}
\usepackage[pdftex, pdftitle={Article}, pdfauthor={Author}]{hyperref}
\hypersetup{colorlinks=true, 
	linkcolor=blue,
	citecolor=orange,
	filecolor=black,
	urlcolor=cyan}

\usepackage{xr}
\newcommand\blfootnote[1]{%
	\begingroup
	\renewcommand\thefootnote{}\footnote{#1}%
	\addtocounter{footnote}{-1}%
	\endgroup
}

\begin{document}

\title{Quantum logic operations and algorithms in a single 25-level atomic qudit}

\author{Pei Jiang Low$^{1,2,\dagger}$}
\author{Nicholas C.F. Zutt$^{1,2,\dagger}$}
\author{Gaurav A. Tathed$^{1,2}$}
\author{Crystal Senko$^{1,2,*}$}

\affiliation{$^1$Institute for Quantum Computing, University of Waterloo, Waterloo, Ontario, N2L 3G1, Canada}
\affiliation{$^2$Department of Physics and Astronomy, University of Waterloo, Waterloo, Ontario, N2L 3G1, Canada}

\begin{abstract}
Scaling quantum computers remains a substantial scientific and technological challenge. 
Leveraging the full range of intrinsic degrees of freedom in quantum systems offers a promising route towards enhanced algorithmic performance and hardware efficiency.
We experimentally study the use of \ce{^{137}Ba^+} ions for quantum information processing, achieving high-fidelity state preparation and readout of up to 25 internal levels, thus forming a 25-dimensional qudit. 
By probing superpositions of up to 24 states, we investigate how errors scale with qudit dimension $d$ and identify the primary error sources affecting quantum coherence. 
Additionally, we demonstrate high-dimensional qudit operations by implementing a 3-qubit Bernstein-Vazirani algorithm and a 4-qubit Toffoli gate with a single ion. 
Our findings suggest that quantum computing architectures based on large-dimensional qudits hold significant promise.
\end{abstract}

\blfootnote{$^\dagger$ These authors contributed equally.}
\blfootnote{$^*$ Corresponding author: \hyperlink{csenko@uwaterloo.ca}{csenko@uwaterloo.ca}.}

\maketitle

\section{Introduction} 
\label{sec:1_introduction}

Efforts to scale quantum systems for fault-tolerant computation have focused on quantum analogues of binary computing~\cite{preskill18_quant_comput_nisq_era_beyon, fedorov22_quant_comput_at_quant_advan_thres, philips22_univer_contr_six_qubit_quant_proces_silic, ai24_quant_error_correc_below_surfac_code_thres}.
However, trapped ions—a leading quantum computing platform~\cite{harty14_high_fidel_prepar_gates_memor, smith24_singl_qubit_gates_with_error_at_level, wright19_bench_quant_comput}—have a richer energy-level structure than the 2 levels typically used for qubits.
Recently, a \textit{qudit} encoding approach, which moves to multi-valued logic for quantum algorithms, has gained traction~\cite{wang20_qudit_high_dimen_quant_comput, chi22_progr_qudit_based_quant_proces, hrmo23_nativ_qudit_entan_trapp_ion_quant_proces, watson15_fast_fault_toler_decod_qubit}.
Qudit-based encoding enables several useful applications, such as: directly working with base $d$ qudits to scale computational space faster per ion~\cite{low20_pract_trapp_ion_protoc_univer, ringbauer22_univer_qudit_quant_proces_with_trapp_ions, campbell12_magic_state_distil_all_prime}, relaxed error-correction thresholds~\cite{omanakuttan24_fault_toler_quant_comput_using, campbell14_enhan_fault_toler_quant_comput_in}, defining multiple virtual qubits per ion~\cite{campbell22_polyq_quant_proces, kiktenko23_realiz_quant_algor_with_qudit, nikolaeva24_effic_realiz_quant_algor_with_qudit, shivam24_utilit_virtual_qubit_trapp_ion_quant_comput}, using \textit{in-situ} measurement-free error correction by encoding qubit states in larger Hilbert spaces~\cite{pirandola08_minim_qudit_code_qubit_phase_dampin_chann, debry25_error_correc_logic_qubit_encod}, and directly simulating higher-dimensional systems~\cite{senko15_realiz_quant_integ_spin_chain, illa24_qu8it_quant_simul_lattic_quant_chrom, meth25_simul_two_dimen_lattic_gauge}.

Trapped ion qudit-based quantum computing has been demonstrated primarily with smaller dimensions ($d \le 7$)~\cite{leupold18_sustain_state_indep_quant_contex, ringbauer22_univer_qudit_quant_proces_with_trapp_ions, aksenov23_realiz_quant_gates_with_optic_addres, hrmo23_nativ_qudit_entan_trapp_ion_quant_proces, zalivako23_contin_dynam_decoup_optic_qudit, zalivako24_towar_multiq_quant_proces_based}.
This constraint is set by the number of (meta)stable states in the chosen ion species.
Increasing $d$ could thus benefit the applications listed above.
For $N$ qudits, computational space scales as $d^N$~\cite{wang20_qudit_high_dimen_quant_comput}.
Bosonic system simulations could benefit from larger truncations of the boson's (infinite) Hilbert space~\cite{meth25_simul_two_dimen_lattic_gauge, yang16_analog_quant_simul_of}.
Extra states enable qubit encodings that correct more known error sources~\cite{gottesman01_encod_qubit_oscil, chiesa20_molec_nanom_as_qubit_with, lim23_fault_toler_qubit_encod_using_spin_qudit}.
These considerations motivate work with isotopes such as \ce{^{137}Ba^+}~\cite{bramman23_ablat_loadin_qudit_measur_barium, low23_contr_reado_high_dimens_trapped}, \ce{^{43}Ca^+}~\cite{benhelm07_measur_hyper_struc}, and \ce{^{173}Yb^+}~\cite{dzuba16_hyper_induc_elect_dipol_contr, allcock21_bluep_trapp_ion_quant_comput}, whose large nuclear spins yield many (meta)stable states.
Laboratory control of this complex energy-level structure, recently used to implement efficient quantum search algorithms~\cite{shi25_effic_implem_quant_algor_with}, will enable computation with up to tens of states per ion.

\begin{figure*}
\centering
\includegraphics[width=0.97\textwidth]{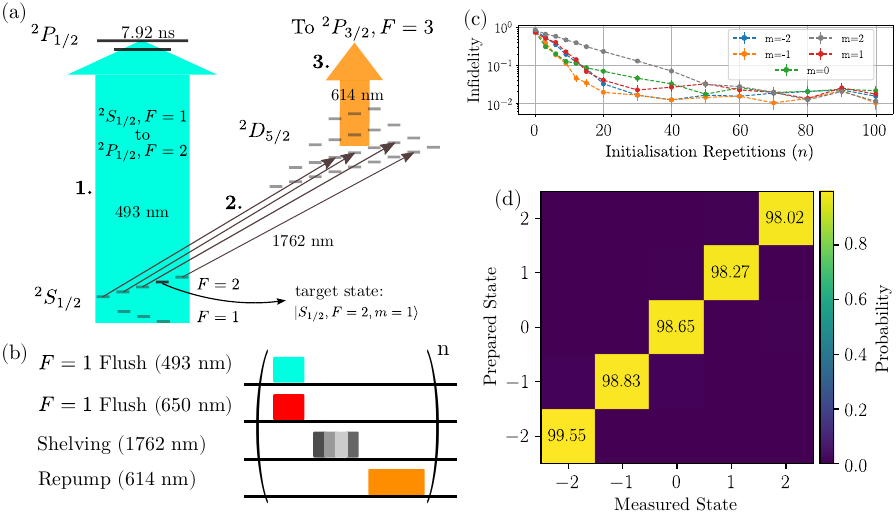}
  \caption[Narrow-band optical pumping (NBOP) initialisation.]{\textbf{Narrow-band optical pumping (NBOP) initialisation.} \textbf{(a)} The ion starts out in one of the 8 states in $S_{1/2}$. 
  The NBOP sequence, consisting of \textbf{1.} $F=1$ state flushing, \textbf{2.} \SI{1762}{\nano\meter} shelving pulses, and \textbf{3.} repumping with \SI{614}{\nano\meter} light, is applied repeatedly, increasing the probability that the initial state falls into the one $S_{1/2}, F=2$ state not being actively pumped out (in this case $\ket{m=1}$).
  \textbf{(b)} Pulse sequence for NBOP. 
  \textbf{(c)} Initialisation infidelities for the 5 states in $S_{1/2}, F=2$ as a function of NBOP sequence repetitions.
  \textbf{(d)} State preparation probabilities at $n=80$ NBOP repetitions. 
  We measure on average \SI{98.6(8)}{\%} combined state preparation and measurement success for $S_{1/2}$ initialisation.
  1000 shots are taken for all data points in \textbf{(c, d)}; error bars denote $1\sigma$ confidence using the Wilson interval~\cite{wilson27_probab_infer_law_succes_statis_infer}.}
  \label{fig:s12_initialisation_with_SPAM}
\end{figure*}

In this work, we demonstrate coherent control and readout of a 25-level \ce{^{137}Ba^{+}} trapped ion qudit encoded in the $S_{1/2}$ and $D_{5/2}$ electronic levels—the maximum number of states discriminable in a single-shot measurement using known protocols.
This is the largest digital encoding of a trapped ion qudit reported in the literature to date.
We show heralded state preparation and measurement with a fidelity of \SI{99.51(5)}{\%}, limited primarily by spontaneous decay of the $D_{5/2}$ states and off-resonant driving.
Leveraging quadrupole transitions that connect all states in the $S_{1/2}$ and $D_{5/2}$ manifolds, we demonstrate coherence between qudit states up to $d=24$ using a Ramsey-type experiment.
We implement the Bernstein-Vazirani algorithm on 2 and 3 virtual qubits in a single \ce{^{137}Ba^+} ion and present a full physical error model—each error source validated in independent experiments—which matches results to within \SI{6.8}{\%} experimental uncertainty.
Our model projects that combined errors from these sources can be reduced to the $10^{-3}$ level or better for $d \leq 16$ using engineering improvements already demonstrated with trapped-ions.

\section{Results}

\begin{figure*}
\centering
\includegraphics[width=0.97\textwidth]{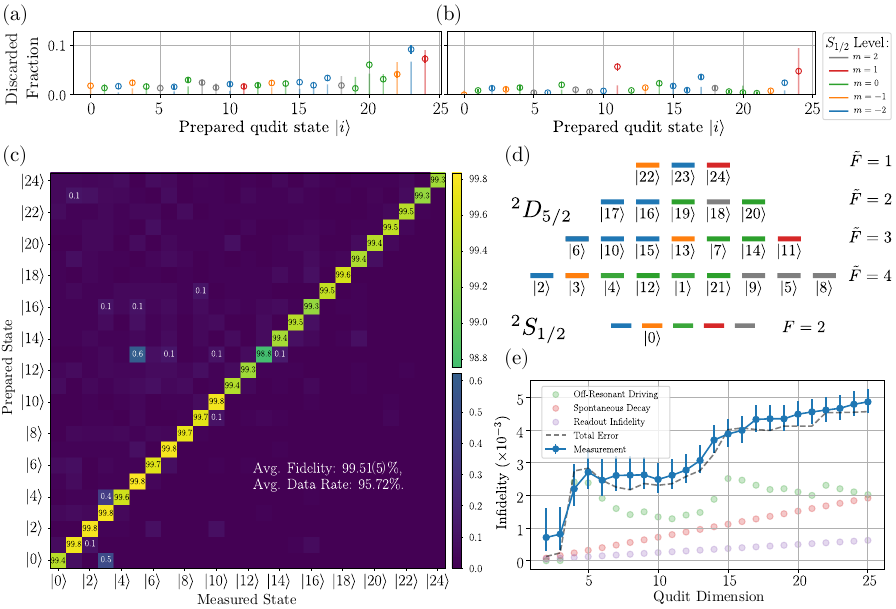}
  \caption{\textbf{25-level SPAM infidelities and data rates.} 
  Data loss rates per qudit basis state $\ket{i}$ due to initialisation errors \textbf{(a)} and measurement errors \textbf{(b)}.
  Open circles are data, bars are theoretical predictions based on transition strengths and magnetic field sensitivities as well as $S_{1/2}$ initialisation error rates.
  \textbf{(c)} SPAM measurement results showing an average fidelity of $\SI{99.51(5)}{\%}$ over all states. 
  Off-diagonal populations above \SI{0.1}{\%} are indicated with text.
  \textbf{(d)} State encoding scheme for SPAM, with colours indicating the $S_{1/2}$ level from which initial states in $D_{5/2}$ are shelved (same colours as legend in \textbf{(a, b)}).
  \textbf{(e.)} Error modelling based on expected off-resonant driving, spontaneous decay from $D_{5/2} \rightarrow S_{1/2}$, and bright/dark discrimination error, per qudit dimension. 
  5000 shots are taken for all data points; error bars denote $1\sigma$ confidence using the Wilson interval~\cite{wilson27_probab_infer_law_succes_statis_infer}.
  }
  \label{fig:d52_SPAM}
\end{figure*}

All results are measured with single \ce{^{137}Ba^+} ions confined in a hand-assembled Paul trap~\cite{white22_isotop_selec_laser_ablat_ion} with secular motional frequencies of (1.27, 1.46, 0.21) MHz.
Doppler cooling and fluorescence detection use \SI{493}{\nano\meter} and \SI{650}{\nano\meter} lasers driving the $S_{1/2}-P_{1/2}$ and $D_{3/2}-P_{1/2}$ transitions, respectively.
A linewidth-narrowed \SI{1762}{\nano\meter} \enquote{shelving} laser, incident with wavevector orthogonal to the weak trap axis, has its frequency tuned via arbitrary voltage waveforms applied to an electro-optic phase modulator.
Its polarization is set so that frequency adjustments alone can drive any quadrupole-allowed transition between the $S_{1/2}$ and $D_{5/2}$ states.
Two (or 5) reference transitions are used to calibrate all \enquote{shelving} transition frequencies (or strengths) (Methods~\ref{sec:A1_experimental_setup}).
A \SI{614}{\nano\meter} laser repumps the $D_{5/2}$ states via $P_{3/2}$ to the $S_{1/2}$ manifold.

\subsection{Qudit encoding and SPAM} 
\label{sec:2_encoding_SPAM}

Qudits are encoded in the $S_{1/2}$ and $D_{5/2}$ magnetic sublevels~\cite{low25_contr_readout_trapp_ion_qudit}.
To extend the Hilbert space to 25 levels, we developed a novel extension to the narrowband optical pumping (NBOP) technique of Ref.~\cite{an22_high_fidel_state_prepar_measur} (see also~\cite{ransford21_weak_dissip_high_fidel_qubit}), enabling preparation of any $S_{1/2}, F=2$ sublevel on demand.

This NBOP protocol, diagrammed in Fig.~\hyperref[fig:s12_initialisation_with_SPAM]{\ref*{fig:s12_initialisation_with_SPAM}a} and~\hyperref[fig:s12_initialisation_with_SPAM]{\ref*{fig:s12_initialisation_with_SPAM}b}, is performed similarly for each target state in $S_{1/2}, F=2$.
The sequence consists of:
(1) flushing the $S_{1/2}, F=1$ levels with \SI{493}{\nano\meter} and \SI{650}{\nano\meter} lasers;
(2) pulsing four frequencies on the \SI{1762}{\nano\meter} shelving laser to empty all $S_{1/2}, F=2$ levels except the target state;
(3) repumping the $D_{5/2}$ states with \SI{614}{\nano\meter} light (tuned to preferentially excite $P_{3/2}, F=3$ states to minimise decay into $S_{1/2}, F=1$) and \SI{650}{\nano\meter} light.
In step (2), the \SI{1762}{\nano\meter} transitions are selected to optimise the optical pumping rate and final population in the targeted $D_{5/2}$ states (Methods~\ref{sec:A2_NBOP}).
This sequence is repeated until the targeted $S_{1/2}$ state population saturates (Fig.~\hyperref[fig:s12_initialisation_with_SPAM]{\ref*{fig:s12_initialisation_with_SPAM}c}).

To measure initialisation fidelity, we probe each initialised state separately.
Six $\pi$-pulses of \SI{1762}{\nano\meter} light shelve population from the initialised $S_{1/2}, F=2$ state to distinct $D_{5/2}$ states, followed by a fluorescence check with \SI{493}{\nano\meter} and \SI{650}{\nano\meter} light.
A \textit{dark} outcome indicates the ion is in $D_{5/2}$ and corresponds to successful initialisation.
This technique mitigates shelving pulse infidelity as a contributor to $S_{1/2}$ SPAM error, at the expense of measuring only one basis state per shot.
We measure a \SI{98.6(8)}{\%} initialisation probability, averaged over all five states, at $n=80$ sequence repetitions (Fig.~\hyperref[fig:s12_initialisation_with_SPAM]{\ref*{fig:s12_initialisation_with_SPAM}d}).
Part of the observed error may arise from off-resonant \SI{493}{\nano\meter} light driving population out of the target state.
We also find evidence that off-resonant \SI{1762}{\nano\meter} excitation to $D_{5/2}$, for example via motional sidebands (Supp. Materials II), contributes to infidelity.

The $D_{5/2}$ state initialisation begins with NBOP to a chosen
$S_{1/2}$ level, usually the state with the strongest transition to the target $D_{5/2}$ state.
A single 1762 nm laser shelving pulse is then used to drive the ion to the desired state. 
We detect initialisation faults with a heralding method similar to Ref.~\cite{sotirova24_high_fidel_heral_quant_state_prepar_measur}, discarding data points with known failures (Fig. 2a).
Following the shelving pulse, a fluorescence check of $S_{1/2}$ results in a bright outcome for unsuccessful initialisation
To employ this method for $S_{1/2}$ state preparation, we shelve the $S_{1/2}$ state to $D_{5/2}$ using a \SI{1762}{\nano\meter} transition with low $\pi$-pulse error. 
We then perform the heralding measurement, keep or discard the computation attempt accordingly, then de-shelve the population from $D_{5/2}$ to the desired $S_{1/2}$ state.

We implement single-shot state measurement of the 25-level system by first checking for fluorescence in the encoded $S_{1/2}$ state, then sequentially de-shelving each $D_{5/2}$ state to $S_{1/2}$ and checking for fluorescence after each pulse.
The first bright outcome is assigned as the measurement result.
If no bright outcome is detected over the 25 levels, this indicates failure to de-shelve from $D_{5/2}$, raising an error flag on that computation~\cite{low25_contr_readout_trapp_ion_qudit, kang23_quant_error_correc_with_metas}, resulting in data loss (Fig.~\hyperref[fig:d52_SPAM]{\ref*{fig:d52_SPAM}b}).
We report combined data loss rates of \SI{2.82(5)}{\%} from initialisation and \SI{1.51(3)}{\%} from measurement.
These heralding techniques enable the \SI{99.51(5)}{\%} SPAM fidelity measured here when averaged over all 25 levels (Fig.~\hyperref[fig:d52_SPAM]{\ref*{fig:d52_SPAM}c}).

Three dominant error sources impact this SPAM result: off-resonant driving, spontaneous decay from $D_{5/2} \rightarrow S_{1/2}$, and bright/dark discrimination error (fluorescence check infidelity).
Due to the two heralding techniques used, the measured SPAM fidelity is robust against \SI{1762}{\nano\meter} laser pulse errors, which are recast as data losses.
The exception is $\ket{0}$, where the initialisation herald is followed by a de-shelving pulse from the $\ket{3}$ encoded state (see Fig.~\hyperref[fig:d52_SPAM]{\ref*{fig:d52_SPAM}d}).
The $\ket{3}\rightarrow\ket{0}$ pulse infidelity introduces a \SI{0.5}{\%} chance of misdiagnosing $\ket{0}$ as $\ket{3}$.
To verify the error model and study qudit SPAM error rates for $d\leq25$ (Fig.~\hyperref[fig:d52_SPAM]{\ref*{fig:d52_SPAM}c}), we re-analyse the data from Fig.~\hyperref[fig:d52_SPAM]{\ref*{fig:d52_SPAM}b}, including only the first $d$ prepared states and reinterpreting heralding outcomes.
The modelled errors agree with experimental data to within measurement precision (Fig.~\hyperref[fig:d52_SPAM]{\ref*{fig:d52_SPAM}e}) and exhibit linear or sub-linear scaling with $d$ (details in Supp. Materials III).

This qudit SPAM measurement is competitive with qubit SPAM results on other platforms~\cite{chen23_trans_qubit_readout_fidel_at, wang24_fidel_measur_super_qubit, blumoff22_fast_high_fidel_state_prepar, serrano24_improv_singl_shot_qubit_readout}, and each error source can be reduced by at least an order of magnitude through optimising magnetic fields, trap geometry, and photon collection efficiency (Supp. Materials V).
We project based on this model that heralded SPAM errors can be engineered to below $10^{-4}$ for qudits with $d\leq 25$, comparable to the best recent SPAM results in any qubit systems~\cite{an22_high_fidel_state_prepar_measur, sotirova24_high_fidel_heral_quant_state_prepar_measur}.

\subsection{Multi-level coherence} 
\label{sec:3_coherent_qudit}

\begin{figure*}
\centering
\includegraphics[width=0.97\textwidth]{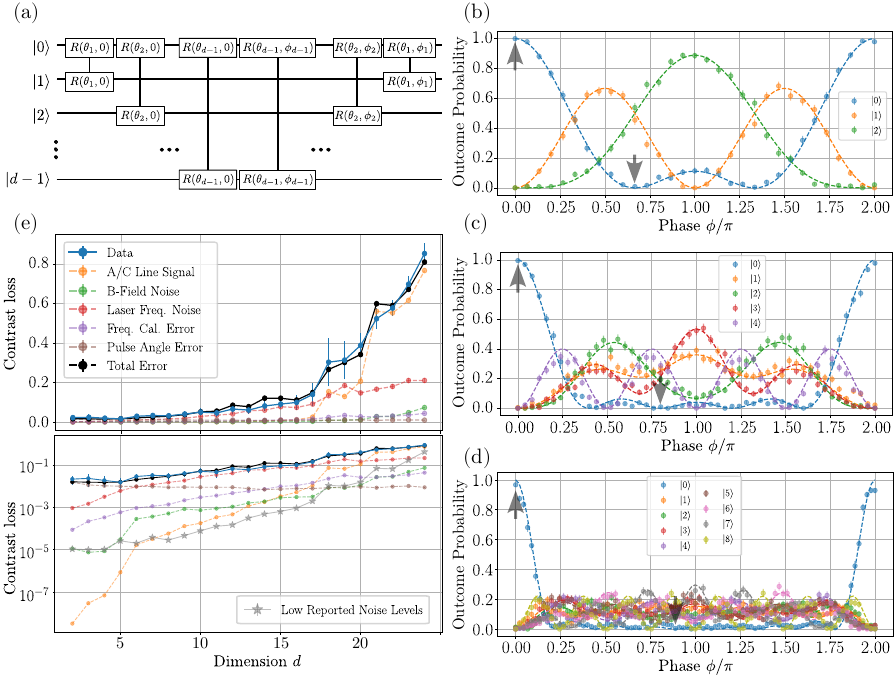} 
\caption[Multi-level coherence probes and dimensional contrast scaling.]{\textbf{Multi-level coherence probes and dimensional contrast scaling.}
\textbf{(a)} Pulse sequence circuit diagram - each rotation addresses a transition between state $\ket{0}$ and $\ket{j}$, written as $R(\theta,\phi)$ for pulse angle $\theta$ and phase $\phi$. $\theta_j = 2\arcsin\left( 1/\sqrt{d-j+1} \right)$, and $\phi_j = \phi \cdot j$ for $\phi$ ranging from 0 to $2\pi$ over the course of a full phase scan.
Multi-level superpositions and coherence probing for \textbf{(b)} $d=3$, \textbf{(c)} $d=5$, and \textbf{(d)} $d=9$, as a function of $\phi$. 
Arrows indicate the phases $\phi$ used to compute the contrast shown in \textbf{(e)}.
Dashed lines are ideal (noiseless) traces as guides for the expected evolution (see Methods~\ref{sec:A4_qudit_Ramseys}). 
\textbf{(e)} Contrasts of the $\ket{0}$ state populations as a function of qudit dimension $d$, along with the prediction based on known noise sources affecting the ion. 
The sub-plots represent the same data on a linear scale (top) and log scale (bottom).
The log scale plot includes simulated qudit Ramsey results with lower noise values reported in literature.
See Methods~\ref{sec:A3_transition_state_choice} for physical state choices and transitions used for each dimension of the qudit Ramsey measurements.
Error bars denote $1\sigma$ confidence using the Wilson interval~\cite{wilson27_probab_infer_law_succes_statis_infer}.}
\label{fig:qudit_Ramsey_phase_scans_contrasts}
\end{figure*}

To demonstrate coherent control of this multi-level system, we developed a protocol generalising qubit Ramsey interferometry techniques, similar to Refs.~\cite{godfrin18_gener_ramsey_inter_explor_with, thenuwara23_ramsey_inter_three_level_five} in other platforms, to probe the mutual coherence of all $d$ qudit states simultaneously.
We study superpositions of up to $d=24$ states.

For each dimension $d$, we select a different subset of physical states to encode the qudit.
For $d < 18$, we optimise a cost function accounting for transition strengths and mutual magnetic-field sensitivities in sets of size $d$ (Methods~\ref{sec:A3_transition_state_choice}).
This optimisation is constrained to a \enquote{star} transition graph with one $S_{1/2}$ state as the central node.
Beyond this, no more $D_{5/2}$ states connect to a single $S_{1/2}$ level with usable transitions, so we extend the $d=17$ encoding by adding $D_{5/2}$ states that minimise the total probe sequence duration.

The protocol prepares an equal superposition of $d$ basis states spanning the qudit, then probes how well the population can be re-phased into a single basis state.
As sketched in Fig.~\hyperref[fig:qudit_Ramsey_phase_scans_contrasts]{\ref*{fig:qudit_Ramsey_phase_scans_contrasts}a}, after initialising to $\ket{0}$, we apply quadrupole transition pulses with rotation angles
\begin{equation}
\theta_j = 2 \arcsin\left( \frac{1}{\sqrt{d-j+1}} \right)
\end{equation}
resonant with transitions to each $\ket{j \neq 0}$.
Here, $\theta_j =\pi$ fully transfers population from $\ket{0}$ to $\ket{j}$.
For $d \leq 17$, the pulse sequence requires only $d-1$ unique transitions; for larger $d$, some states are populated via multi-step transitions through intermediate $S_{1/2}$ and $D_{5/2}$ states not in the encoding.
A second pulse sequence drives the same transitions for the same durations, in reverse order, applying phases $\phi_j = \phi \cdot j$ with $\phi$ a freely varying parameter.
Ideal and experimental outcomes are plotted versus $\phi$ (Figs.~\hyperref[fig:qudit_Ramsey_phase_scans_contrasts]{\ref*{fig:qudit_Ramsey_phase_scans_contrasts}b-d}).

In an ideal noise-free scenario, when $\phi$ is an integer multiple of $2\pi/d$, the $\ket{0}$ population at the end of the protocol is either \SI{100}{\%} if $\phi$ is a multiple of $2\pi$, or \SI{0}{\%} otherwise (derivation in Methods~\ref{sec:A4_qudit_Ramseys}).
To obtain a global figure of merit for mutual coherence of the qudit system, we measure $\ket{0}$ population at $\phi=0$ and at $\phi=\pi$ for even $d$ (or $\pi(d-1)/d$ for odd $d$).
The difference, or contrast, is \SI{100}{\%} in the ideal case, while decoherence and control pulse imperfections reduce this value.

Contrast measurements as a function of qudit dimension $d$ (Fig.~\hyperref[fig:qudit_Ramsey_phase_scans_contrasts]{\ref*{fig:qudit_Ramsey_phase_scans_contrasts}e}) show how multipartite superpositions are coherently manipulated in this system.
As more states are added, contrast decreases due to inclusion of states with differing magnetic field sensitivities and longer pulse sequences.
A sharp drop in contrast occurs for $d>17$, where lengthier multi-pulse rotations are required.
Independent measurements (Supp. Materials IV) identify the primary error sources as magnetic field noise, laser frequency and power fluctuations, transition frequency and strength miscalibration, and A/C line-induced magnetic field changes, which dephase transitions.
Phase coherence between an $S_{1/2}$ state and a $D_{5/2}$ state is sensitive to both laser frequency and magnetic field noise, while coherence between two states in the same manifold is unaffected by laser frequency noise, even when using multi-step \SI{1762}{\nano\meter} transitions (Supp. Materials IV).
We measure coherence times within $D_{5/2}$ up to two orders of magnitude longer than between the $S_{1/2}$ and $D_{5/2}$ manifolds.

We combine independent error measurements into a full physical error model with no free parameters, using shot-to-shot Monte Carlo sampling (Methods~\ref{sec:A5_noise_simulations}).
The model yields results within experimental precision of the measured qudit contrasts (solid black line in Fig.~\hyperref[fig:qudit_Ramsey_phase_scans_contrasts]{\ref*{fig:qudit_Ramsey_phase_scans_contrasts}e}).
This modelling identifies the dominant error sources for each qudit dimension $d$: pulse angle error dominates at lower $d$, while laser frequency noise and A/C line-induced magnetic field changes dominate at higher $d$.
Our current trap apparatus lacks fibre noise cancellation and magnetic field shielding, both of which would mitigate these dominant errors.

Using literature-reported noise levels, our model projects contrast loss below $10^{-4}$ with demonstrated technologies for $d \leq 10$ (star markers in Fig.~\hyperref[fig:qudit_Ramsey_phase_scans_contrasts]{\ref*{fig:qudit_Ramsey_phase_scans_contrasts}e}).
Even at $d=16$, we predict a $1.0(2) \times 10^{-3}$ contrast loss for these multi-level superpositions.
These estimates assume \SI{0.04}{\micro G} magnetic field fluctuations, \SI{70}{\micro G} A/C line signal amplitude~\cite{ruster16_long_lived_zeeman_trapp_ion_qubit}, \SI{0.5}{\hertz} laser linewidths~\cite{jiang08_long_distan_frequen_trans_over, alnis08_subher_linew_diode_laser_by}, \SI{10}{Hz} frequency calibration errors enabled by longer coherence times (Supp. Materials I), and \SI{0.1}{\%} pulse angle errors.
In Supp. Materials V, we detail the engineering specifications required to reduce these modelled errors to $10^{-4}$ or better.
While this pulse sequence may not reflect a full universal gate set, it indicates the errors achievable for single-qudit unitary operations manipulating all $d$ states coherently.

\subsection{Quantum processing with virtual qubits}
\label{sec:4_BVA_algorithm}

\begin{figure*}
\centering
\includegraphics[width=0.97\textwidth]{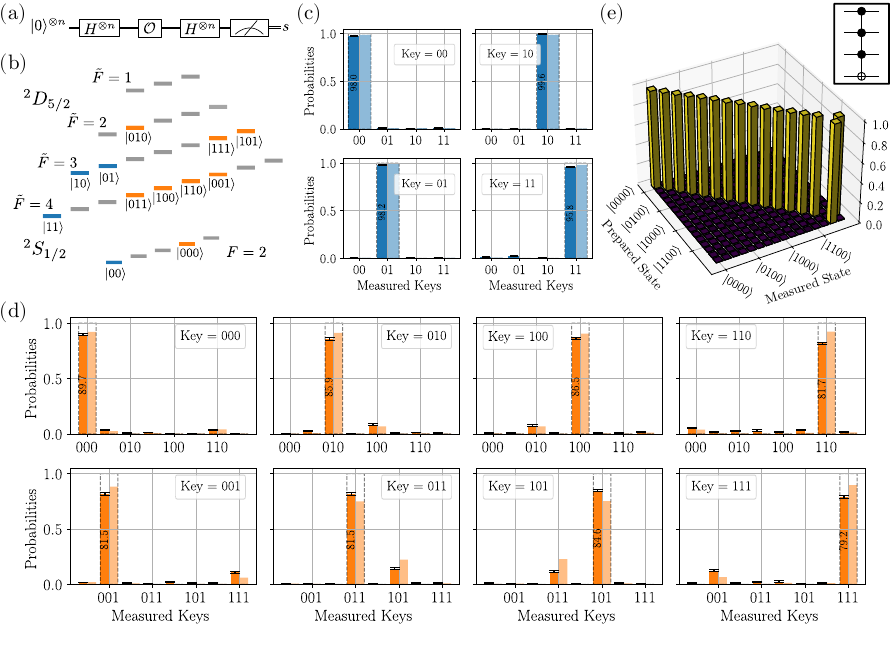}
\caption{\textbf{Bernstein-Vazirani circuit diagram, state selection, 2 and 3 virtual qubit results, and \textit{CCCNOT} truth table.} 
\textbf{(a)} Bernstein-Vazirani algorithm circuit diagram.
\textbf{(b)} State choices for 2 virtual qubit (blue) and 3 virtual qubit (orange) implementations of the key finding algorithm (Methods~\ref{sec:A3_transition_state_choice}).
\textbf{(c)} 2-bit secret key finding succeeds with a probability of \SI{97.9(2)}{\%}, with the simulated results (lighter blue bars) predicted at \SI{98.6}{\%} success. 
\textbf{(d)} 3-bit secret key finding results in a \SI{83.8(8)}{\%} probability of success. The simulated result (light orange bars) with noise modelling predicts the "$XXI$" type errors seen here, and shows an \SI{86.9}{\%} success probability on average. See Methods~\ref{sec:A6_unitary_decomposition} for pulse finding method, Supp. Materials VI for full pulse sequences used.
\textbf{(e)} Truth table measurement of a $CCCNOT$ gate using a four virtual qubit encoding. On average, the correct output state is achieved with a probability of \SI{99.5(2)}{\%}, inclusive of SPAM error. (Inset) Circuit diagram illustrating three virtual qubits acting as controls, with one target qubit.
1000 shots are taken for data in (c-e), and error bars denote $1\sigma$ confidence using the Wilson interval~\cite{wilson27_probab_infer_law_succes_statis_infer}.}
\label{fig:virtqubit_BVA_results}
\end{figure*}

Though qudits of any dimension $d \leq 25$ can be encoded, we demonstrate quantum gates and circuits in the specific scenario of encoding $n$ virtual qubits using $d=2^n$ basis states.
We select an optimized subset of states (Methods~\ref{sec:A3_transition_state_choice}) and implement full control using the single \SI{1762}{\nano\meter} phase- and frequency-controlled laser to drive quadrupole transitions between $S_{1/2} \leftrightarrow D_{5/2}$.
This enables decomposition of any unitary operation (single- and multi-qubit gates) into sequences of Givens rotations between pairs of states~\cite{brennen05_criter_exact_qudit_univer, brylinski02_mathem_quant_compu}.

We implement the Bernstein-Vazirani key finding algorithm~\cite{bernstein97_quant_compl_theor}, given as a circuit diagram in Fig.~\hyperref[fig:virtqubit_BVA_results]{\ref*{fig:virtqubit_BVA_results}a}.
The algorithm aims to determine an unknown $n$-bit key $s$, given an oracle implementing $f: {0,1}^n \rightarrow {0,1}$, where $f(x)$ is $x \cdot s$ modulo 2 for secret $s \in {0,1}^n$.
First, an equal superposition of all $2^n$ basis states is prepared via $n$ parallel Hadamard gates, $H^{\otimes n}$, bringing the system from
\begin{equation}
\ket{0}^{\otimes n} \rightarrow \frac{1}{\sqrt{2^n}} \sum\limits_{x=0}^{2^n-1}\ket{x}
\end{equation}
for computational basis states $\ket{x}$.
The oracle $\mathcal{O}$ transforms this state via
\begin{equation}
\ket{x} \rightarrow (-1)^{f(x)}\ket{x}.
\end{equation}
A second $H^{\otimes n}$ causes all population to coherently recombine in $\ket{s}$, allowing measurement to reveal $s$.
This algorithm demonstrates quantum advantage by requiring a single oracle call, compared to $n$ calls classically required.

Initialising the system in $\ket{0}^{\otimes n}$ for $n \in {2,3}$ amounts to preparing the ion in the appropriate $\ket{m_f}$ state in $S_{1/2}$ (encodings in Fig.~\hyperref[fig:virtqubit_BVA_results]{\ref*{fig:virtqubit_BVA_results}b}).
We find pulse decompositions implementing $H^{\otimes n}$ using 5 Givens rotations for two virtual qubits and 21 for three (Supp. Materials VI).
The oracle $\ket{x} \rightarrow (-1)^{f(x)}\ket{x}$ is implemented via virtual-Z gates, by adding $\pi$ to the phase of rotations involving basis states $\ket{x}$ in the second $H^{\otimes n}$.
This ensures the oracle is noiseless and instantaneous.

To minimise algorithm run-time for three virtual qubits, we abbreviate the initial $H^{\otimes n}$ gate before the oracle by exploiting the fact that the system always starts in $\ket{0}^n$.
This allows direct evolution to $\frac{1}{\sqrt{2^n}}\sum_0^{2^n-1}\ket{x}$ using $2^n-1$ rotations.
A full $H^{\otimes 3}$ gate with \SI{1.2}{\milli\second} total pulse time is thus replaced by a \enquote{superposition pulse} of 7 rotations in \SI{170}{\micro\second} (Supp. Materials VI).

This implementation achieves a successful key-finding probability of \SI{97.9(2)}{\%} for 2 virtual qubits and \SI{83.8(8)}{\%} for 3 virtual qubits (Figs.~\hyperref[fig:virtqubit_BVA_results]{\ref*{fig:virtqubit_BVA_results}c},~\hyperref[fig:virtqubit_BVA_results]{\ref*{fig:virtqubit_BVA_results}d}).
These results are on par with recent multi-qubit implementations of the same algorithm~\cite{fallek16_trans_implem_berns_vazir_algor}.
We attribute the comparable fidelities in part to the lack of multi-ion gate errors, highlighting an advantage of the virtual qubit approach.
Ordinarily, Hadamard gates $H^{\otimes n}$ could be applied in parallel across separate qubits; in this system, they are an \textit{unfavourable} gate type for virtual qubit encoding.
We are currently limited to sequential Givens rotations, resulting in longer run times compared to parallel gates on separate ions, though this could be improved by simultaneous driving of transitions.
No active noise mitigation techniques are used apart from triggering experiment start times to the A/C line signal.
Using our error model, we find that A/C line magnetic field changes, despite triggering, dominate the error, causing the $XXI$-type errors in the 3 virtual qubit result (Fig.~\hyperref[fig:virtqubit_BVA_results]{\ref*{fig:virtqubit_BVA_results}d}).
Based on literature-reported noise levels (Supp. Materials V), we predict a mean failure probability of $4.0(3)\times 10^{-3}$ for 3-bit key finding is attainable with current technology.

Unlike the relatively intensive $H^{\otimes n}$ gates, complex gates such as Toffoli or $CCCNOT$ can be implemented via a swap between $\ket{1110}$ and $\ket{1111}$ in the 4 virtual qubit encoding (Fig.~\hyperref[fig:virtqubit_BVA_results]{\ref*{fig:virtqubit_BVA_results}e}).
The $CCCNOT$ gate achieves a truth-table fidelity of \SI{99.5(2)}{\%} averaged over the 16 basis states, limited mainly by SPAM errors, outperforming similar gates on multiple physical qubits~\cite{figgatt17_compl_grover_searc_progr_quant_comput, nikolaeva24_scalab_improv_gener_toffol_gate}.
Trade-offs between quantum encoding strategies (qubit, qudit, or multiple virtual qubits) remain an active area of study~\cite{keppens25_qudit_vs_qubit_simul_perform, wang20_qudit_high_dimen_quant_comput} and are vital for assessing multi-level computing advantages.
Our results highlight the importance of qudit state selection and pulse decomposition in evaluating these trade-offs.

\section{Discussion} 
\label{sec:5_outlook}

We have demonstrated most of the primitives required for quantum information processing~\cite{divincenzo00_physic_implem_quant_comput}: SPAM, multi-level coherence, and single-ion control.
The high-fidelity SPAM and coherent control demonstrated above establish the viability of high-dimensional qudit encoding in \ce{^{137}Ba^+}.
The pulse decomposition scheme used for the Bernstein-Vazirani algorithm can be extended to derive a full single-qudit gate set, as the phase and frequency control of the quadrupole \SI{1762}{\nano\meter} laser enables arbitrary unitary evolutions~\cite{brennen05_criter_exact_qudit_univer, brennen05_effic_circuit_exact_univer_comput_with_qudit}.

Developing optimal control sequences for this super-dense trapped ion encoding may reveal additional practical trade-offs.
The advantage of requiring fewer two-ion gates must be balanced against the increased complexity of single-qubit gates~\cite{shivam24_utilit_virtual_qubit_trapp_ion_quant_comput} to determine the most useful applications of this scheme.
Universal computation is possible through both native qudit entanglement~\cite{hrmo23_nativ_qudit_entan_trapp_ion_quant_proces} and standard qubit M{\o}lmer-S{\o}rensen gates~\cite{brennen05_effic_circuit_exact_univer_comput_with_qudit, ringbauer22_univer_qudit_quant_proces_with_trapp_ions}, recently demonstrated in \ce{^{137}Ba^+}~\cite{wang25_exper_realiz_direc_entan_gates}.

Our results open a new region of the trade-off space in quantum processor architectural design.
This hardware platform can motivate research into optimal quantum compiling strategies for high-dimensional qudits and up to 4 virtual qubits per atom, as well as algorithm co-design with hardware-aware qudit state selection and gate decomposition.
Whether used as additional data states or ancillary resources, the extra states native to trapped ion systems offer multiple roles for advancing quantum computation.

\section*{Acknowledgements} 
\label{sec:acknowledgements}

This research was supported, in part, by the Natural Sciences and Engineering Research Council of Canada
(NSERC), RGPIN-2018-05253 and the Canada First Research Excellence Fund (CFREF) (Transformative Quantum
Technologies), CFREF-2015-00011. C.S. is also supported by a Canada Research Chair.

\textbf{Author Contributions:} P.J.L., G.A.T. and N.C.F.Z. conducted the experiments. C.S. supervised the project. All authors conceived experiments, analysed the results, and reviewed the manuscript.

\textbf{Competing Interests:} The authors declare no competing interests.

\textbf{Data and code availability:} The data underlying this work, and the code used for data analysis and presentation, are publicly available at~\cite{low25_25level_qudit_zenodo_dataset}.

\bibliographystyle{apsrev4-1}
\bibliography{references}

\clearpage

\appendix*
\renewcommand{\appendixname}{}
\section*{METHODS}

\subsection{Experimental setup}
\label{sec:A1_experimental_setup}

We use the same experimental setup as outlined in our previous work~\cite{low25_contr_readout_trapp_ion_qudit}, with some changes to the experimental parameters.
We employ a two-step, isotope selective ablation loading scheme as detailed in~\cite{white22_isotop_selec_laser_ablat_ion}.
In this work, the laser intensities are \SI{13}{\milli\watt/\milli\meter^2} for \SI{493}{\nano\meter}, \SI{113}{\milli\watt/\milli\meter^2} for \SI{650}{\nano\meter}, and \SI{1.68}{\milli\watt/\milli\meter^2} for \SI{614}{\nano\meter}.
The applied magnetic field strength is $\sim$\SI{4.209}{G}. 
The magnetic field is supplied by an array of nickel-plated neodymium permanent magnets (\ce{Nd_2 Fe_{14} B}) held in 3D printed mounts fixed directly to the trap vacuum chamber.

Unless otherwise stated, all coherent operations initiated after state preparation is completed are triggered on the rising edge of the laboratory's alternating current (A/C) power.
This mitigates fluctuations in magnetic field from shot-to-shot that stem from the \SI{60}{\hertz} A/C that powers all lab equipment, a well-known and documented effect in trapped ion setups~\cite{benhelm08_towar_fault_toler_quant_comput, hu23_compen_power_line_induc_magnet}.

System calibration proceeds as follows (details in Supp. Materials I):
Frequencies used to drive the quadrupole $S_{1/2} \leftrightarrow D_{5/2}$ transitions are calibrated by measuring two reference transitions which enable calibration of magnetic field drifts as well as slow laser cavity drifts.
Frequency calibration is typically run every 5-10 minutes during data acquisition.
Transition strengths are calibrated by measuring the Rabi frequencies of five reference transitions, each with a unique $\Delta m \in \{-2, -1, 0, 1, 2\}$. 
These serve as a reference for every other transition with the matching $\Delta m$ value, which are calculated from theory and scaled up/down according to the output of the reference measurement. 
These Rabi frequencies are typically calibrated once per day prior to data acquisition.

\subsection{Narrow-band optical pumping}
\label{sec:A2_NBOP}

The NBOP approach introduced in Sec.~\ref{sec:2_encoding_SPAM} involves several steps which each have opportunity for tuning/optimisation.
Firstly, the $S_{1/2}, F=1$ states flushing step is sensitive to the power and pulse time of the \SI{493}{\nano\meter} light used.
A lower power, short pulse is desirable to reduce off-resonant excitation of the $S_{1/2}, F=2$ states, as exciting these states would scramble the intended initial state.
In the experiments shown in this work, the \SI{493}{\nano\meter} light has an intensity of $\sim \SI{0.5}{\milli\watt/\milli\meter^2}$, and the flushing step has a duration of \SI{1.5}{\micro\second}.

In addition, the transitions chosen for NBOP initialisation, which vary depending on which of the 5 $S_{1/2}, F=2$ levels is initialised, must be chosen with care.
We choose to repump population from $D_{5/2}$ preferentially through $P_{3/2}, F=3$ such that decay from $P_{3/2}$ to $S_{1/2},F=1$ is suppressed by dipole selection rules, which also helps to reduce the number of NBOP cycles required before initialisation fidelity saturates.
With this in mind, we have calculated the re-pumping path probabilities for all states in $D_{5/2}$ through $^2P_{/2}, F=3$ and decaying to $S_{1/2}, F=2$ in order to choose which transitions have the best chance of moving population toward the intended initial state efficiently.
These repump path probabilities (shown in Supp. Materials II) have a large impact on the initialisation efficiency and fidelity.
To give just one example, if one intends to initialise $\ket{F=2, m=-2}$ in $S_{1/2}$, choosing to drive 
population in $\ket{F=2, m=2}$ to the state $\ket{D_{5/2}, \tilde{F}=4, m_{\tilde{F}}=4}$ would not work at all, as this state can \textit{only} be pumped to $\ket{P_{3/2}, F=3, m=3}$ which in turn can \textit{only} decay back to $\ket{S_{1/2},F=2, m=2}$ (neglecting multi-step processes like repeated driving to $P_{3/2}$ following decay back to $D_{5/2}$ or $D_{3/2}$).
This would form a closed loop preventing any population from getting to the desired initial state.
Beyond completely forbidden transition paths, one must choose repump paths with care, as weak pathways to moving population to the desired initial state will unduly increase the number of repetitions needed.
Given that our fluorescent readout completely scrambles the initial state of the next experiment, and therefore returns a mixed state of all levels in $S_{1/2}$, this would severely limit the overall initialisation fidelity.

\subsection{State choices for encoding SPAM, qudits, and virtual qubits}
\label{sec:A3_transition_state_choice}

\centerline{\textit{\textbf{State preparation and measurement}}}
\vspace{1em}

For the $d=25$ level SPAM result, all states in $D_{5/2}$ are encoded, along with a single state in $S_{1/2}, F=2$. 
The $\ket{0}$ state in $S_{1/2}$ was determined to be the state with the shortest transition $\pi$-time from any state in $D_{5/2}$. 
This ensures that, after shelving the population to $D_{5/2}$ for the initialisation herald, the de-shelving pulse uses a strong transition and the infidelity of that $\pi$-pulse population transfer is minimised. 
The $D_{5/2}$ that is used for the initialisation herald for $\ket{0}$ is the $\ket{3}$ state, which accounts for the population found in $\ket{3}$ for the prepared $\ket{0}$ outcomes in Fig.~\hyperref[fig:d52_SPAM]{\ref*{fig:d52_SPAM}d}. 
The encoded states $\ket{1} \rightarrow \ket{24}$ in $D_{5/2}$ are labelled by the readout order, where the readout order is mainly determined by the de-shelving pulse transition strength, from highest to lowest. 
This ensures that the shortest de-shelving pulses are used earliest, to minimise off-resonant de-shelving of population in other $D_{5/2}$ states during the sequential readout.
Some changes to that ordering scheme were subsequently made to account for and mitigate off-resonant effects found empirically.

\vspace{1em}
\centerline{\textit{\textbf{Multi-level encoding}}}
\vspace{1em}

Multi-level encoding for coherent operations follows a different method since the mutual coherence and connectivity of many states are the most important qualities to optimise over.
We begin by first using the information we have on all transition strengths to determine the fastest transition times between any pair of states in the joint manifolds of $\{D_{5/2}, S_{1/2}, F=2\}$ using a fully exhaustive search of all transition paths between states.
This gives us every pair-wise effective transition $\pi$-time, $\tau_{\pi,(j,k)}$, between any states $\ket{j},\ket{k}$. Then, for a set of states $\{S\}$ we calculate the total of all transition times 
\begin{equation}
    \tau_{\{S\}} = \sum\limits_{j<k} \tau_{\pi,(j,k)}.
\end{equation}

In order to evaluate a cost function to optimise over and find an appropriate set of states, we need some kind of coherence information on this group of states as well. To that end, we use measured magnetic field noise, as well as laser phase noise (Supp. Materials IV), and the magnetic field dependence of the energy levels themselves (Supp. Fig.~\ref{suppfig:calibration_overview}), to calculate the pairwise expected coherence due to three separate sources for a transition between two states $j$ and $k$: Gaussian laser phase/frequency noise $T_{L,G}^{j,k}$, Lorentzian laser phase/frequency noise $T_{L,L}^{j,k}$, and Gaussian magnetic field noise $T_{B,G}^{j,k}$.

For a set of states $\{S\}$, we then define the cost function as

\begin{equation}
    \mathcal{C}(\{S\}) = \frac{1}{l} \sum\limits_{j<k} \frac{\tau_{\{S\}}^2}{(T_{L,G}^{j,k})^2} + \sum\limits_{j<k} \frac{\tau_{\{S\}}}{T_{L,L}^{j,k}} + \sum\limits_{j<k} \frac{\tau_{\{S\}}^2}{(T_{B,G}^{j,k})^2}
\end{equation}
where $l$ is the number of states in the set. $l$ may be equal to the qudit dimension $d$, or greater, if all states in the set are not directly connected (i.e. a mediating state, which remains un-encoded).

Summing over all pairs of states in a given group, regardless of whether they are directly connected or not, and including mediating states, has the effect of introducing a rather steep penalty for state sets that do not have all participating states encoded (that is, where $l > d$).
Nonetheless, for $d \le 17$, we impose a final restriction on the state sets that all states must be connected to the same, single state in $S_{1/2}, F=2$, a so-called \enquote{star-topology} graph, or tree with one node and $d - 1$ leaves.
With these constraints, and this cost function, we exhaustively search all possible sets of states of size $d$ and pick the set that minimises $\mathcal{C}$.

This connectivity graph enables the generation of equal superposition states, as in the qudit Ramsey-type measurements shown in Sec.~\ref{sec:3_coherent_qudit}, with one pulse per participating state $\ket{j \neq 0}$, with the label $\ket{0}$ assigned to the state in $S_{1/2}, F=2$.

Beyond $d = 17$, a star-topology is no longer practical, as the transitions from any possible initial $S_{1/2}, F=2$ state to the remaining states in $D_{5/2}$ are either too weak and magnetic field sensitive to be reliably used (see Supp. Materials I) or completely forbidden by quadrupole transition selection rules.
As a result, some multi-pulse transitions are required in order to generate superpositions for $d \ge 18$.
The state sets used for $d \ge 18$ include the states used for $d = 17$ (where $\ket{0} = \ket{S_{1/2}, F=2, m=0}$ state), with additional states found manually.

The 4, 8, and 16 state encodings used for basis states in the Bernstein-Vazirani algorithm, as well as the $CCCNOT$ gate implementation, largely follow the state set finding results above.
The two exceptions to this are states encoded for the $CCCNOT$ gate which are weakly connected to the $\ket{0}=\ket{S_{1/2}, F=2, m=0}$ state but allow higher data rates using initialisation and measurement via transitions to other $S_{1/2}, F=2$ states. See Supp. Materials VI for details on all level encodings.

\subsection{Qudit Ramsey-type experiment - analytic expressions}
\label{sec:A4_qudit_Ramseys}

In intermediate magnetic field strengths of order \SI{1}{G}, in the context of maximally encoding the $D_{5/2}$ states in \ce{^{137}Ba^+}, it is more practical to utilise the \SI{1762}{\nano \meter} for coherent manipulation and connecting the qudit states as compared to driving the $D_{5/2}$ to $D_{5/2}$ transitions directly, i.e. with stimulated Raman transitions.
This is because the closest frequency separation between all $D_{5/2}$ to $D_{5/2}$ transitions is in the order of \SI{1}{\kilo \hertz} in magnetic field strengths of order \SI{1}{G}, which would limit the gate speed significantly to minimise off-resonant transition error.
From our previous work, we have calculated that there are sets of $S_{1/2}$ to $D_{5/2}$ quadrupole transitions with sufficiently strong oscillator strengths to connect any two states in the $S_{1/2}$ and $D_{5/2}$ subspace~\cite{low25_contr_readout_trapp_ion_qudit}.
Thus, in this work, we use the \SI{1762}{\nano \meter} quadrupole transitions for complete coherent control of a \ce{^{137}Ba^+} qudit.

To demonstrate simultaneous coherent control of multiple qudit states, we perform a qudit Ramsey-type experiment, which we describe as follows.
From a prepared encoded energy eigenstate, which we encode as $\ket{0}$, a series of \SI{1762}{\nano \meter} laser pulses prepares an equal superposition of the $d$ encoded states.
Each \SI{1762}{\nano \meter} laser pulse drives a two-level transition to bring a $\frac{1}{d - l + 1}$ fraction of the state population in $\ket{0}$ to $\ket{l}$, described by the unitary
\begin{align}
    \begin{split}
    \hat{U}_{1,l} &= \sqrt{\frac{d-l}{d+1-l}} \left( \ketbra{0}{0} + \ketbra{l}{l} \right) \\
    &+ \sqrt{\frac{1}{d+1-l}} \left( \ketbra{l}{0} - \ketbra{0}{l} \right) + \sum_{m \ne 0,l} \ketbra{m}{m}.
    \end{split}
    \label{eq:U1_qudit_unitary}
\end{align}
The pulse sequence is ordered to distribute state population from $\ket{0}$ to $\ket{l}$ in the sequence from $l=1$ to $l=d-1$, described by the unitary
\begin{equation}
    \hat{U}_{1} = \hat{U}_{1,d-1}\hat{U}_{1,d-2}\cdots\hat{U}_{1,2}\hat{U}_{1,1}
\end{equation}
and it can be verified that
\begin{equation}
    \hat{U}_{1} \ket{0} = \sum_{l=0}^{d-1} \sqrt{\frac{1}{d}} \ket{l}.
\end{equation}
A second pulse sequence performing the following unitary is then sent to the qudit ion.
\begin{equation}
    \hat{U}_{2} = \hat{U}_{2,1}\hat{U}_{2,2}\cdots\hat{U}_{2,d-2}\hat{U}_{2,d-1}
\end{equation}

\begin{align}
    \begin{split}
    \hat{U}_{2,l} &= \sqrt{\frac{d-l}{d+1-l}} \left( \ketbra{0}{0} + \ketbra{l}{l} \right) \\ 
    &+ \sqrt{\frac{1}{d+1-l}} \left( - e^{-i l \phi} \ketbra{l}{0} + e^{i l \phi} \ketbra{0}{l} \right) \\ 
    &+ \sum_{m \ne 0,l} \ketbra{m}{m}.
    \end{split}
    \label{eq:U2_qudit_unitary}
\end{align}
where $\phi$ is a variable phase.
It can then be derived that the state population for each state after the pulse sequences is
\begin{align}
    \begin{split}
    P_l &= \lvert \langle l \lvert \hat{U}_{2} \hat{U}_{1} \rvert 0 \rangle \rvert^2 \\ 
    &= \begin{cases} 
    \frac{1}{d} + \frac{2}{d^2} \sum_{m=1}^{d-1} \left( d - m \right) \cos \left( m \phi \right), & l = 0 \\ 
    \frac{1}{d} - \frac{1}{d} \cos \left( \left( d - 1 \right) \phi \right), & l = d-1 \\ 
    g(l,d,\phi), & otherwise 
    \end{cases}
    \label{eq:Qudit_Ramsey_perfect_function}
    \end{split}
\end{align}
where
\begin{align}
    \begin{split}
    & g(l,d,\phi) = \frac{1}{d} + \frac{2}{d \left( d-l+1 \right) \left( d-l \right)} \sum_{m=l+1}^{d-1} \cos \left( m \phi \right)  \\
    &- \frac{2}{d(d-l+1)} \left( \cos \left( l \phi \right) + \sum_{m=1}^{d-1-l} \left( \frac{m+1}{d-l} \cos \left( m \phi \right) \right) \right)   
    \end{split}
\end{align}
From Eq.~\ref{eq:Qudit_Ramsey_perfect_function}, we can see that we get a perfect recovery of the $\ket{0}$ state population when $\phi$ is zero or an integer multiple of $2 \pi$. 
At phases equal to integer multiples of $\frac{2 \pi}{d}$, 
the $\ket{0}$ state population is zero, except at $\phi=0,2\pi n$.
With a time delay of $t$ between $\hat{U}_1$ and $\hat{U}_2$, and some small frequency detunings of $\Delta_l$ ($\Delta_l \ll \Omega_l$ where $\Omega_l$ is the Rabi frequency for the $\ket{0}$ to $\ket{l}$ transition), the $\ket{0}$ state population is
\begin{align}
    \begin{split}
    P_0 &= \frac{1}{d} + \frac{2}{d^2} \sum_{m=1}^{d-1} \cos \left( m \phi + \Delta_m t \right) \\
    &+ \frac{2}{d^2} \sum_{n=1}^{d-2} \sum_{m=l+1}^{d-1} \cos \left( \left( m - l \right) \phi + \left( \Delta_m - \Delta_l \right) t \right).
    \label{eq:Qudit_Ramsey_detuning_P0_function}
    \end{split}
\end{align}
From Eq.~\ref{eq:Qudit_Ramsey_detuning_P0_function}, the first $\frac{1}{d}$ term is the equilibrium state population if there is no coherence, the second set of terms relate to coherence of $\ket{0}$ state and other non-zero states, the third set of terms relate to the coherence between non-zero states.
It can be seen that with large dephasing errors (i.e. large and randomly distributed $\Delta_n t$), the cosine terms interfere destructively and $P_0$ stays at approximately $\frac{1}{d}$ at all $\phi$.
Thus, comparing the contrast of $P_0$ at different values of $\phi$ is a good metric for characterizing the coherence of the qudit, with 1 being perfectly coherent and 0 when completely decohered.

\subsection{Monte-Carlo simulation of noisy multi-level system}
\label{sec:A5_noise_simulations}

Here, we explain how the simulated results for contrast measurements shown in Fig.~\hyperref[fig:qudit_Ramsey_phase_scans_contrasts]{\ref*{fig:qudit_Ramsey_phase_scans_contrasts}d}, as well as simulated Bernstein-Vazirani algorithm results shown in Fig.~\ref{fig:virtqubit_BVA_results}, are calculated.

For a given system with $d$ dimensions, we begin by defining the unitaries associated with each pulse in the qudit Ramsey-type pulse sequence, or in the Bernstein-Vazirani algorithm.
That is, each Givens rotation between states $\ket{0}$ and $\ket{m}$ is constructed from a Hamiltonian of the form

\begin{equation}
    \hat{H}_p^{(0,m)} = \hspace{0.5em}
    \begin{blockarray}{ccccc}
     \ket{0} &  & \ket{m} & & & \\
    \begin{block}{(ccccc)}
    0 & ... & \frac{\Omega}{2}e^{i\phi} & ... & 0 \\
    \vdots & \ddots & & &  \\
    \frac{\Omega}{2}e^{-i\phi} & & \Delta\omega & & \vdots \\
    \vdots & & & \ddots &  \\
    0 & & ...  & & 0 \\
    \end{block}
    \end{blockarray}
    \label{eq:pulses_hamiltonian}
\end{equation}
where $\Omega$ is the Rabi frequency of the quadrupole transition from $\ket{0} \leftrightarrow \ket{m}$, $\Delta\omega$ is the detuning between the transition's true frequency at that moment in time and the laser's frequency for driving the transition in question, and $\phi$ is the phase accumulated due to the detuning integrated over time since the start of the experimental shot.
The constraint that all transitions are driven by quadrupole transitions means that, for each Hamiltonian of the form in Eq.~\ref{eq:pulses_hamiltonian}, $\ket{0}$ is in $S_{1/2}$ and $\ket{m}$ is in $D_{5/2}$. 

Several noise sources contribute to the pulse's detuning $\Delta\omega$ and changes to the applied phase $\phi$, where each noise source is constant over a single experiment, with shot-to-shot variability being sampled according to distributions whose profiles are informed by past measurements (see Supp. Materials IV).

\begin{table}
    \vspace{2em}
    \centering
    \begin{tabular}{|c|c|c|c|}
    \hline
        Noise Type & Symbol & Profile & Width \\
        \hline
         Magnetic Field & $\delta_B$ & Gaussian & $\SI{24}{\micro G}$ \\
         Laser Frequency & $\delta_L$ & Voigt & \SI{287}{\hertz} \\
         Frequency Miscalibration & $\delta_f$ & Gaussian & \SI{296}{\hertz} \\
         Pulse-Time Miscalibration & $\delta_\tau$ & Gaussian & \SI{1.77}{\%}  \\
         Pulse-Time Drift & $\delta_p$ & Gaussian & \SI{2.61}{\%}  \\
         \hline
    \end{tabular}
    \caption{\textbf{Noise sources, profiles, and distribution widths.} Above we show the various contributions to detuning and pulse-time errors in the Monte Carlo simulations of qudit Ramsey-type experiments and the Bernstein-Vazirani algorithm. Further discussion on the measurements leading to these profiles and widths can be found in Supp. Materials IV. The Voigt profile width reported here is calculated from the decay of a magnetically insensitive transition, and followed the formula from \cite{kielkopf73_new_approx_to_voigt_funct} for deriving a width from Gaussian and Lorentzian components.}
    \label{tab:noise_sources_distributions}
\end{table}

Without noise, $\Delta\omega$ for each pulse Hamiltonian $\hat{H}_p^{(n,m)}$ is 0, and the phases $\phi = \phi_{0,m} +\Delta \phi_{m}$ are just $\phi_{0,m}$. 
That is, all are set according to those phases required for the implementation of unitaries $\hat{U}_{1,l}$ and $\hat{U}_{2,l}$ as in Eqs.~\ref{eq:U1_qudit_unitary} and~\ref{eq:U2_qudit_unitary}.
With noise turned on, a single Monte-Carlo sample proceeds by first determining the values for the noise sources identified in Table~\ref{tab:noise_sources_distributions}; each is sampled from a distribution with the indicated profile and width.
One more contribution to the detuning $\Delta\omega$ and the phase $\phi$ comes from the measured dependence of the magnetic field value with respect to the \SI{60}{\hertz} A/C wall power signal. This contribution has been measured and shown to vary according to

\begin{multline}
\delta_{B,\text{line}}(t) = A_{60} \sin(\omega_{60} t + \phi_{60}) \\
+ A_{180} \sin(\omega_{180} t + \phi_{180})
\label{eq:line_signal_expression}
\end{multline}
where $A_{60}=\SI{128}{\micro G}$, $A_{180}=\SI{40}{\micro G}$, $\phi_{60}=\SI{-0.636}{rad}$, and $\phi_{60}=\SI{-1.551}{rad}$ (see measurement in Supp. Fig.~\ref{suppfig:ac_line_signal}). 
This noise source is unique in the sense that it is reproducible, and therefore it is not a contribution that is sampled from shot-to-shot but rather introduces predictable offsets to the accumulated phases $\phi$ and the transition detunings $\Delta\omega$.
(It should also, in principle, be possible to compensate for this well-defined signal - though this has not been implemented in the results shown here.)

From the values sampled for the magnetic field noise $\delta_B$, the laser frequency fluctuation $\delta_L$, the frequency calibration error $\delta_f$, and the pulse time error $\delta_\tau$ for a given shot, the detuning $\Delta\omega$ for a given pulse $\hat{H}_p^{(0,m)}$ can be calculated to be

\begin{equation}
    \Delta\omega = 2\pi \cdot \left( \kappa_{0,m} \left(\delta_B + \delta_{B,line}(t) \right) + \delta_L + \delta_f \right)
\end{equation}
where $\kappa_{0,m}$ is the sensitivity (in MHz/G) of the transition driven in $\hat{H}_p^{(0,m)}$, and $t$ is the time at which the pulse begins relative to the (line-triggered) start of the experiment. 

The accumulated phase $\phi$ on a given transition is then essentially an integral of the detuning above for the time leading up to the application of the pulse, and can be expressed as

\begin{multline}
\Delta\phi_m = 2\pi t \cdot \left( \kappa_{0,m}\delta_B + \delta_L + \delta_f \right) \\
+ 2\pi \kappa_{0,m} \int_0^t \delta_{B,\text{line}}(t') \, dt'
\label{eq:delta_phi_expression}
\end{multline}
where the form of this expression makes clear the fact that we assume a constant $\delta_{B/L/f}$ for each Monte Carlo sample. This value $\Delta\phi_m$ is added to the ideal phase of the pulse.

Imperfections due to laser intensity fluctuations are modelled by sampling the measured errors in transition $\pi$-time predictions from calibration error $\delta\tau_c$, as well as the measured drift in the pulse times $\delta\tau_d$ over time scales similar to experimental time scales and altering the transition Rabi frequency to be 

\begin{equation}
    \Omega \rightarrow \frac{\Omega}{1+\delta\tau_t+\delta\tau_p}.
\end{equation}
Again, $\delta\tau_{c/d}$ is assumed constant over the course of an experimental run, and is sampled once per Monte Carlo run.

Noise-less simulations of the qudit Ramsey-like pulse sequences yield idealised evolutions, as shown with dashed lines in figs.~\hyperref[fig:qudit_Ramsey_phase_scans_contrasts]{\ref*{fig:qudit_Ramsey_phase_scans_contrasts}a-c}. In~\hyperref[fig:qudit_Ramsey_phase_scans_contrasts]{\ref*{fig:qudit_Ramsey_phase_scans_contrasts}d}, we run the Monte Carlo simulation with 1024 shots in total, with each noise source individually (dashed lines) and then with all noise sources together (solid black line) in order to compare with the experimental results. Simulated results in~\hyperref[fig:virtqubit_BVA_results]{\ref*{fig:virtqubit_BVA_results}b,d} also use 1024 Monte-Carlo samples.

\subsection{Unitary decomposition with Star-Topology constraint}
\label{sec:A6_unitary_decomposition}

Based on state encoding choices (Methods~\ref{sec:A3_transition_state_choice}) we have one computational state in $S_{1/2}$, $|0\rangle\equiv\ket{F=2,\,m_F}_{\mathrm{6S}_{1/2}}$, while the $\mathrm{6D}_{5/2}$ manifold supplies the remaining $d-1$ states of qudit (or virtual qubit) register, $\{\ket{1},\dots,\ket{d-1}\}\equiv
\{\ket{F,\,m_F}\}_{\mathrm{5D}_{5/2}}$. Coherent operations are driven by a set of \SI{1762}{\nano\meter} electric-quadrupole transitions that couple $\ket{0}$ in $\mathrm{^2S}_{1/2}$ to all the states in $\mathrm{^2D}_{5/2}$.
Hence, the resulting interaction graph has a \emph{star topology} with a central node at $\ket{0}$.

The restricted connectivity invalidates standard QR-decomposition protocols, which assume full pairwise Givens rotations. We modify the algorithm so that all rotations act in a two-dimensional subspace $ \mathrm{span} \bigl \{|0\rangle,\ket{i}\bigr\}$ ($i \in \{1, \dots, d-1\}$), ensuring physical implementability.  

\vspace{1em}
\centerline{\textit{\textbf{Decomposition Strategy}}}
\vspace{1em}

Let $U$ be the target unitary expressed in the logical basis $\{\ket{0}, \ket{1},\dots, \ket{d-1}\}$. We define Givens rotations acting on indices $0$ and $i$ $(i>1)$ as 
\begin{equation}
\label{eq:star_givens}
\begin{aligned}
&G_{0,i}(\theta) \; =\; 
      \exp\!\Bigl[-\theta\bigl(
        \ketbra{0}{i}-\ketbra{i}{0}\bigr)\Bigr] \\[6pt]
 &=\;
      \bordermatrix{
            & \ket{0} &                  & \ket{i} &                   \cr
            &  \cos\theta  &      0        &  \sin\theta  &0              \cr
            &     0         & \mathbb{I}_{i-2} & 0           &0              \cr
            & -\sin\theta  &  0            &  \cos\theta  &0              \cr
            &0              & 0                 &0            & \mathbb{I}_{d-1-i}\cr
      }_{\!\!\!0,i}
\end{aligned}
\end{equation}

Our objective is to rewrite an arbitrary unitary $U$ as a time-ordered list of physically allowed Givens pulses. 

\begin{align}
U &= G_{0}^{\dagger}\,G_{1}^{\dagger}\dotsm G_{N-1}^{\dagger}, \\
G_{j} &=G_{0,i_j}(\theta_j),\ 
i_j\in\{1,\dots,d-1\},
\end{align}
so that playing the pulses \(\{G_{N-1},\dots,G_{0}\}\) on the ion implements \(U\) exactly within the star connectivity. The algorithm starts from $V_{0}=U$ and pre-multiplies
by Givens rotations until the running matrix is the identity,
$V_{N}=I$. Each step preserves one pivot element while annihilating a single off-diagonal entry. We implement the following steps:

\begin{enumerate}
\item \textbf{Initial sweep.}  
      Clear the first column by applying
      $G_{0,i}\bigl(\theta_{i}^{(0)} \bigr)$
      for $i=\{1,2,\dots,d-2,d-1\}$ in some order $\mathcal{O}$.

\begin{equation}
\theta_{i}^{(0)}
\;=\;
\begin{cases}
 \tan^{-1}\!\left(\dfrac{V_{i0}}{V_{00}}\right), & V_{00}>0,\\[6pt]
 \tan^{-1}\!\left(\dfrac{V_{i0}}{V_{00}}\right)+\pi, & V_{00}<0,\\[6pt]
 \tfrac{\pi}{2}\,\operatorname{sgn}(V_{i0}), & V_{00}=0,
\end{cases}
\end{equation}
      
      When finished, column~0 equals the basis vector \(\ket{0}\).

\item \textbf{Swap cycle} (for columns $k=1,\dots,d-1$ in reversed order $\mathcal{O}$).  
      \begin{enumerate}
      \item \emph{Swap.}  
            A fixed \(\pi/2\) pulse $S_{0k}=G_{0,k}(\pi/2)$ exchanges rows $0$ and $k$, moving the would-be diagonal entry into the pivot position
            $V_{0k}$.
      \item \emph{Column elimination.}  
            For each row $i>k$ (processed in order $\mathcal{O}$)
            apply $G_{0,i}(\theta_{ik})$
            to zero $V_{ik}$ while leaving the pivot $V_{0k}$ unchanged.
      \item \emph{Swap-back.}  
            Re-apply $S_{0k}$ to restore the original row ordering.
      \end{enumerate}
\end{enumerate}

Repeating the swap cycle for every column drives \(V\) to
the identity: $G_{N-1}\dotsm G_{0}\,U = I$. Therefore, reversing the list and taking adjoints yields the pulse
sequence for \(U\). The routine uses
$$N \;=\; (d-1)\;+\;\sum_{k=1}^{d-1}\bigl[\,2 + (d-1-k)\bigr]
      \;=\;\frac{(d-1)(d+4)}{2}$$
Givens rotations.

\begin{algorithm}[H]
  \caption{Unitary decomposition (Star topology)}
  \label{alg:star_qr}
  \begin{algorithmic}[1]
    \Require{$U\in U(d)$,\; ordering $\mathcal{O}=(1,\dots,d-1)$,\;
             tolerance $\varepsilon = 10^{-12}$}
    \State $V \gets U$
    \State $\mathcal{R}\gets[\;]$ \Comment{rotation record}
    \Statex
    \Comment{\textbf{--- Initial sweep clears column 0 ---}}
    \For{$i \in \mathrm{reverse}(\mathcal{O})$}
        \If{$|V_{i0}|>\varepsilon$}
            \State $\theta \gets \arctan2(V_{i0},\,V_{00})$
            \State $G \gets G_{0,i}(\theta)$;\; $V \gets G\,V$
            \State $\mathcal{R}.\text{append}(G)$
        \EndIf
    \EndFor
    \Statex
    \Comment{\textbf{--- Swap-cycle for each remaining column ---}}
    \For{$k\in\mathcal{O}$}
        \State $S \gets G_{0,k}(\pi/2)$  \Comment{swap rows \(0,k\)}
        \State $V\gets S\,V$;\; $\mathcal{R}.\text{append}(S)$
        \For{$i \in \mathrm{reverse}(\mathcal{O}[k{:}])$}
            \If{$|V_{ik}|>\varepsilon$}
                \State $\theta \gets \arctan2(V_{ik},\,V_{0k})$
                \State $G \gets G_{0,i}(\theta)$;\; $V\gets G\,V$
                \State $\mathcal{R}.\text{append}(G)$
            \EndIf
        \EndFor
        \State $V\gets S\,V$;\; $\mathcal{R}.\text{append}(S)$
    \EndFor
    \State \Return{$\mathcal{R}$ (sequence of pulses), $V$}
  \end{algorithmic}
\end{algorithm}

Each entry of \(\mathcal{R}\) is a laser pulse on the
\SI{1762}{\nano\meter} quadrupole line.A rotation $G_{0,i}(\theta)$ is effected by a resonant pulse of duration $\tau=\theta/\Omega_{0,i}$, where $\Omega_{0,i}$ is the calibrated Rabi frequency for the $\ket{0}\,\leftrightarrow\ket{i}$ transition.Swap operations $S$ are simply $\pi/2$ pulses on the same link. Concatenating the pulses in $\mathcal{R}$ (time-ordered left to right) implements the desired unitary $U$ exactly within the star topology.

\vspace{1em}
\centerline{\textit{\textbf{Iterative Sequence Compression}}}
\vspace{1em}

While the above method produces a pulse sequence to implement the exact target unitary, it is rarely optimal in practice. The algorithm yields a pulse list $\bigl\{\,i_m,\theta^{(0)}_m,\phi^{(0)}\,\bigr\}_{m=1}^{M}$, here $M$ is total length of sequence. $\theta_m^{(0)}$ is $m^{\text{th}}$ rotation angle. And every pulse carries the same phase $\phi^{(0)}$ (a consequence of the fixed sign convention in~\ref{eq:star_givens}). This uniformity lets us fuse back-to-back rotations that address the same transitions $(0, i)$:
$$R_{0,i}\!\bigl(\pi \theta^{(0)}_{p}\bigr)\,
R_{0,i}\!\bigl(\pi \theta^{(0)}_{p+1}\bigr)
\;=\;
R_{0,i}\!\bigl(\pi\,(\theta^{(0)}_{p}+\theta^{(0)}_{p+1})\bigr),$$
The new sequence will have $M_0(\leq M)$ pulses.

We then treat the two continuous parameters per pulse - angle of rotation $\theta_m \in [0,2]$ and phase $\phi_m \in [0,2]$ - as optimization variables and minimise the Frobenius-norm cost 
$$\mathcal{L}(\mathbf \theta,\boldsymbol\phi)
  \;=\;
  \bigl\lVert
      U_{\mathbf \theta,\boldsymbol\phi}
      \;-\;
      U_{\mathrm{tgt}}
  \bigr\rVert_{F}^{2},$$
using a Limited-memory Broyden-Fletcher-Goldfarb-Shanno Bound (\textsc{L-BFGS-B}) routine that enforces box constraints. Phase updates are implemented experimentally as instantaneous frame-shifts of the local oscillator; they are realised cost-free \emph{virtual $Z$ gates}. 

Finally, we perform a sequential pulse-elimination sweep:
\begin{enumerate}
    \item For each pulse index $m$, temporarily delete  pulse $m$, re-optimise the reduced parameter set, and evaluate the residual error $\mathcal{L}$
    \item if $\mathcal{L} < \epsilon ( \simeq 10^{-3})$ accept the deletion and restart the scan, otherwise keep the pulse.  
\end{enumerate}

The process terminates when no single-pulse deletion keeps the error below $\epsilon$. On average, the process reduces the depth of the pulse sequence by $\sim \SI{35}{\%}$ (based on $\mathcal{O}$), in the case of Hadamard decomposition. The shorter execution window reduces exposure to laser-amplitude noise and magnetic-field drift, offering a practical path toward high-fidelity multi-level gates within the native star topology.

\end{document}


\title{Supplementary Materials for: Quantum logic operations and algorithms in a single 25-level atomic qudit}

\author{Pei Jiang Low$^{1,2,\dagger}$}
\author{Nicholas C.F. Zutt$^{1,2,\dagger}$}
\author{Gaurav A. Tathed$^{1,2}$}
\author{Crystal Senko$^{1,2,*}$}

\affiliation{$^1$Institute for Quantum Computing, University of Waterloo, Waterloo, Ontario, N2L 3G1, Canada}
\affiliation{$^2$Department of Physics and Astronomy, University of Waterloo, Waterloo, Ontario, N2L 3G1, Canada}

\blfootnote{$^\dagger$ These authors contributed equally.}
\blfootnote{$^*$ Corresponding author: \hyperlink{csenko@uwaterloo.ca}{csenko@uwaterloo.ca}.}

\maketitle

\section{System Hamiltonian, Frequencies, and Transition Strengths} 
\label{sec:1_transition_frequencies_strengths}

Here, we discuss the energy level structure of $\ce{^{137}Ba^+}$, particularly the $D_{5/2}$ manifold. 
A detailed understanding of the structure is crucial to developing the control that enables the use of all states in this dense manifold for quantum information processing.

\subsection{System Hamiltonian}
\label{sec:system_hamiltonian}

The Hamiltonian we employ for the unpaired valence electron in $\ce{^{137}Ba^+}$ includes terms for the fine structure splitting, hyperfine structure, and Zeeman splitting (at non-zero magnetic fields). 
In the $S_{1/2}$ manifold, any contributions to energy levels due to fine structure affect all eigenstates equally (such as the Darwin term), and therefore to solve the internal level splittings it suffices to consider just the hyperfine and Zeeman terms. 
We can take a similar approach for the $D_{5/2}$ manifold (following \cite{low25_contr_readout_trapp_ion_qudit} but adding dependence on the nuclear magnetic octupole moment) by noting that the fine-structure splitting of the $5D_{3/2}$ and $D_{5/2}$ levels of $\ce{^{137}Ba^+}$ is $\sim \SI{24}{\tera\hertz}$~\cite{curry04_compil_wavel_energ_level_trans}, which is much larger than any contributions from the hyperfine or Zeeman terms within $D_{5/2}$, so we can obtain accurate energy level calculations by considering just the two terms

\begin{equation}
  \label{eq:summed_hamiltonian}
  \hat{H} = \hat{H}_{HF} + \hat{H}_Z.
\end{equation}

The hyperfine Hamiltonian can be expressed, up to the octupole term and in units 
of $\hbar =1$, as~\cite{steck07_quantum_atom_optic}

\begin{multline}
  \label{eq:hf_hamiltonian}
  \hat{H}_{HF} = A_D \mathbf{I}\cdot\mathbf{J} + B_Q \frac{3(\mathbf{I}\cdot\mathbf{J})^2
    + \frac{3}{2}(\mathbf{I}\cdot\mathbf{J})-I(I+1)J(J+1)}{2I(2I-1)J(2J-1)} \\
  + C_O \frac{10(\mathbf{I}\cdot\mathbf{J})^3 + 20(\mathbf{I}\cdot\mathbf{J})^2
    - 5I(I+1)J(J+1)}{I(I-1)(2I-1)J(J-1)(2J-1)} \\
  + C_O \frac{2(\mathbf{I}\cdot\mathbf{J})\left[ I(I+1) + J(J+1) + 3 - 3I(I+1)J(J+1)
    \right]}{I(I-1)(2I-1)J(J-1)(2J-1)},
\end{multline}
where $A_D$ is the magnetic dipole hyperfine structure constant, $B_Q$ is the electric quadrupole hyperfine structure constant, $C_O$ is the magnetic octupole hyperfine structure constant, $\mathbf{I}$ and $\mathbf{J}$ are the nuclear and electronic angular momentum vectors respectively, and $I$ and $J$ are the nuclear and electronic angular momentum quantum numbers respectively.

The Zeeman term can be expressed as
\cite{steck07_quantum_atom_optic}

\begin{equation}
  \label{eq:zeeman_hamiltonian}
  \hat{H}_Z = \abs{\mathbf{B}} \mu_B (g_J \hat{J}_z + g_I \hat{I}_z),
\end{equation}
where $\mu_B \equiv e\hbar/2m_e$ is the Bohr magneton, $g_I = 0.62490(1) m_e/m_p$ is the nuclear g-factor~\cite{beloy08_nuclear_magnet_octup_momen_hyper_struc}, $g_J$ is the electronic g-factor, and $\hat{I}_z$ and $\hat{J}_z$ are the components of the angular momentum operators along the quantisation axis for the nucleus and electron respectively. 
(We define the $z$-axis to be co-linear with the magnetic field and quantisation axes.)

\begin{figure}[ht!] \centering
  \includegraphics[width=1.0\textwidth]{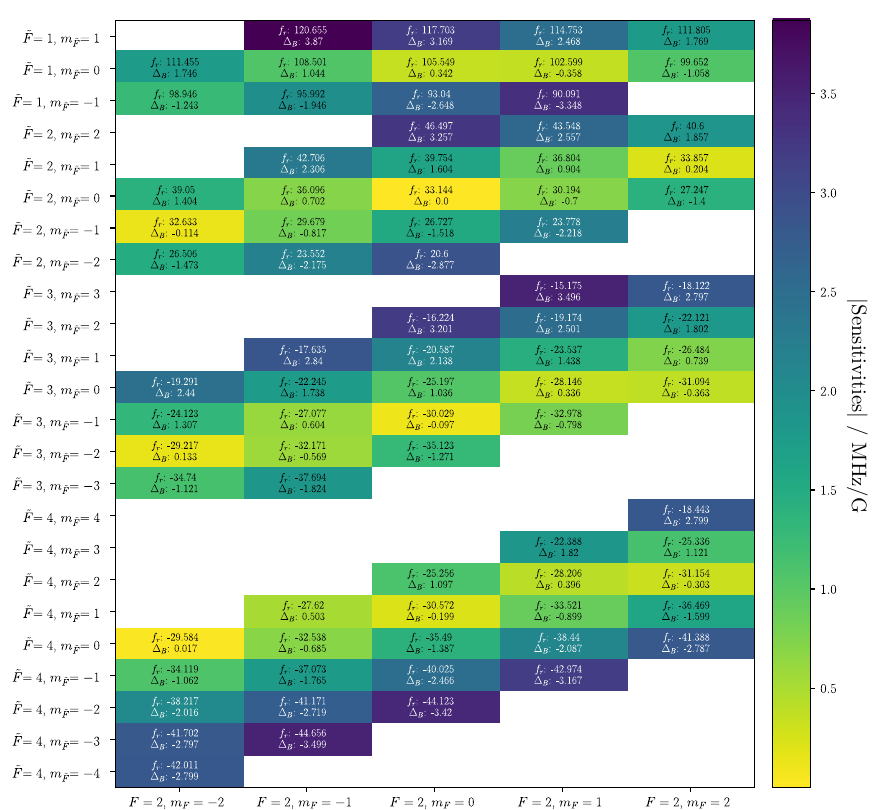}
  \caption[Quadrupole transition frequencies and magnetic field sensitivities.]{\textbf{Quadrupole transition frequencies and magnetic field sensitivities.} 
  Solving the full Hamiltonian in Equation \ref{eq:summed_hamiltonian} allows for the prediction of transition frequencies and the calculation of transition sensitivities to changes in the magnetic field, shown here for a magnetic field of $\abs{\mathbf{B}}=\SI{4.209}{G}$. 
  The colour displays the absolute value of the sensitivities, in order highlight the differences between relatively sensitive and insensitive transitions. 
  The $x$-axis are the $S_{1/2}$ states, the $y$-axis are the $D_{5/2}$ states.
  The frequencies are listed relative to the $S_{1/2} \leftrightarrow D_{5/2}$ transition of \ce{^{138}Ba^+} at 0 field, plus the shift from $S_{1/2}, F=2$ in \ce{^{137}Ba^+}, which is \SI{3014.153159}{\mega\hertz}.}
  \label{suppfig:transition_frequencies_sensitivities}
\end{figure}

\subsection{Transition frequencies}
\label{sec:transition_frequencies}

The transition frequencies predicted by solving the Hamiltonian from Section \ref{sec:system_hamiltonian} can be fit to the experimentally observed frequencies by fitting the spacing of two transitions with differing magnetic field sensitivity to fix the one outlying free parameter in the system - the magnetic field. 
The magnetic field sensitivities of each transition can be directly computed as the sum of the magnetic field dependence of both participating states at a given magnetic field value.
The resulting frequencies and sensitivities are shown in Supp.~Fig.~\ref{suppfig:transition_frequencies_sensitivities} for a magnetic field of $\abs{\mathbf{B}} = \SI{4.209}{G}$. 
The frequencies are listed relative to the $S_{1/2} \leftrightarrow D_{5/2}$ transition of \ce{^{138}Ba^+} at 0 field, plus the shift from $S_{1/2}, F=2$ in \ce{^{137}Ba^+}, which is \SI{3014.153159}{\mega\hertz}.

In practice, we find that solving Equation \ref{eq:summed_hamiltonian} can predict subsequent transition frequencies to within $\sim \SI{2}{\kilo\hertz}$.
This discrepancy likely arises from terms excluded from the Hamiltonian which would account for higher order terms in the hyperfine interaction, or state mixing between the $D_{3/2}$ and $D_{5/2}$ manifolds~\cite{lewty13_exper_deter_nuclear_magnet_octup}. 
In order to find all the relevant $S_{1/2} \leftrightarrow D_{5/2}$ transitions, we employ an efficient calibration scheme which allows us to measure just four data points, corresponding to two different reference transitions transitions, and in so doing calibrate the frequencies of all 80 allowed transitions in about 2 minutes of measurement time. 
The first transition (denoted $f_{int}$) is magnetic field insensitive ($< \SI{1}{\kilo\hertz / G}$): $\ket{S_{1/2}, F=2, m_F=0} \leftrightarrow \ket{D_{5/2}, \tilde{F}=2, m_{\tilde{F}}=0}$. 
The second ($f_{sens}$), with a sensitivity of \SI{3.49}{\mega\hertz /G}, is $|S_{1/2}, F=2, m_F=-1\rangle \leftrightarrow |D_{5/2}, \tilde{F}=4, m_{\tilde{F}}=3 \rangle$. 
These two enable calibration of magnetic field drifts as well as slow laser cavity drifts.

\begin{figure} \centering
  \includegraphics[width=1.0\textwidth]{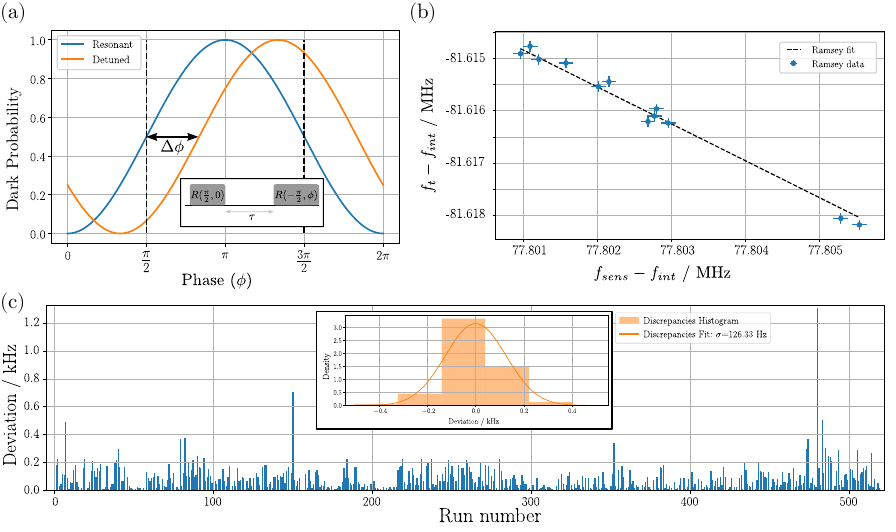}
  \caption[Frequency calibrations overview.]{\textbf{Frequency calibrations overview.} 
  \textbf{(a)} Diagram explaining the Ramsey calibration scheme using just two measurements at a revival pulse phase of $\pi/2$ and $3\pi/2$, with pulse sequence inset. 
  \textbf{(b)} Linear relation on the same transition as in \textbf{(a)}. Each point plotted here represents a triplet measurement of $f_{int},$ $f_{sens}$, and $f_t$. 
  The frequencies of the two reference transitions ($f_{int}$ for the insensitive transition, $f_{sens}$ for the magnetic field sensitive transition) are used along with a target transition ($f_t$) to extract a linear relation plot for each target transition.
  In this plot the data for the transition from $\ket{S_{1/2},F=2,m_F=1} \leftrightarrow \ket{D_{5/2}, \tilde{F}=1, m_{\tilde{F}}=1}$ is shown.
  \textbf{(d)} Deviations between measured transition frequencies and the fit lines, as for example calculated from \textbf{(b)}. 
  (inset) Extracted histogram of deviations from fit predictions with Gaussian fit of standard deviation $\sigma = \SI{126.3}{\hertz}$ shown, giving us the accuracy of this calibration scheme.}
  \label{suppfig:calibration_overview}
\end{figure}

\subsubsection{Ramsey-based calibration scheme}
\label{sec:ramsey_calibration_scheme}

The Ramsey-based calibration scheme necessitates having an already decent estimate for the transition frequency one wishes to calibrate - we therefore use preliminary frequency scan data (data not shown here) in order to gain a rough estimate to begin with.
This initial calibration consists of scanning over the frequency while applying the pulse for the $\pi$-time of the given transition and fitting the resulting Rabi fringe to find the central, resonant frequency.

In order to measure a single target transition frequency $f_t$, we implement a Ramsey type pulse sequence, as shown in the inset of Supp.~Fig.~\hyperref[suppfig:calibration_overview]{\ref*{suppfig:calibration_overview}a}, which creates a superposition of the two states that participate in the target transition using an initial $R(\pi/2,0)$ pulse - where $R(\theta,\phi)$ denotes a rotation of angle $\theta$ about an axis in the $xy$-plane of the Bloch sphere making an angle $\phi$ to the $x$-axis. 
When the laser is perfectly resonant with the two-level transition, no net phase between the laser's rotating frame and that of the ion will accumulate during the time $\tau$ for which the superposition is maintained. 
When applying the reverse pulse, at variable rotation axes using $R(-\pi/2,\phi)$, one would then recover the blue sine curve in Supp.~Fig.~\hyperref[suppfig:calibration_overview]{\ref*{suppfig:calibration_overview}a}.
With any detuning $f_\Delta$ present, though, phase accumulation during the waiting period $\tau$ will result in a revival pulse with a shifted phase, described by $R(-\pi/2,\phi+\Delta)$, yielding a shifted sine curve similar to the orange curve in Supp.~Fig.~\hyperref[suppfig:calibration_overview]{\ref*{suppfig:calibration_overview}a}.
From this shifted phase, one can find the detuning value (details below) that led to this phase offset - but \textit{only assuming} the total phase accumulated is less than $\pm \pi/2$.
This is why a rough calibration of the frequency in question (to within around \SI{1}{\kilo\hertz}) is necessary before starting this routine.

We then use this scheme to build a historical record of how our target transitions' frequencies vary with respect to the two reference transitions $f_{int}$ and $f_{sens}$.
We identify a set of 40 total transitions from Supp.~Fig.~\ref{suppfig:transition_frequencies_sensitivities} which include at least one transition from each $D_{5/2}$ and $S_{1/2}$ state.
For each transition in this set, we measure a triplet consisting of $f_{int}$, $f_{sens}$ and that target transition $f_t$, in succession (this takes around 4 minutes total).
These three values allow us to calculate $f_t - f_{int}$ and $f_{sens}-f_{int}$, and repeated such triplet measurements yield a scatter plot.
Out of the 40 such scatter plots generated in this way, one is shown in Supp.~Fig.~\hyperref[suppfig:calibration_overview]{\ref*{suppfig:calibration_overview}b}, where $f_t$ is $\ket{S_{1/2},F=2,m_F=1} \leftrightarrow \ket{D_{5/2}, \tilde{F}=1, m_{\tilde{F}}=1}$. 

For this scatter plot in Supp.~Fig.~\hyperref[suppfig:calibration_overview]{\ref*{suppfig:calibration_overview}b}, $f_{sens}f_{int}$ is directly proportional to the magnetic field at the time the triplet measurement is taken, and the variation of $f_t-f_{int}$ should therefore depend on the magnetic field sensitivity of the target transition.
The slopes for the data points in these scatter plots are not fit, but rather calculated directly from the magnetic field sensitivities in found from theory, given by

\begin{equation}
  \label{eq:slopes_from_sensitivities}
  m_t = \frac{\Delta_{B,t}-\Delta_{B,int}}{\Delta_{B,sens}-\Delta_{B,int}},
\end{equation}
where $\Delta_{B,i}$ for $i\in\{int, sens, t\}$ denotes the magnetic field sensitivity, in MHz/G, of transition $i$. 
As can be seen from Supp.~Fig.~\hyperref[suppfig:calibration_overview]{\ref*{suppfig:calibration_overview}b}, these imposed slopes match well with the measured calibration data for the Ramsey scheme. 
In order to quantify how well they match, we can compare the expected uncertainty in each individual frequency measurement using the Ramsey scheme with the mean absolute deviation from the fit lines with imposed slopes given from theory. 
In order to make that comparison, we need an estimate for the uncertainty using this scheme, which is calculated in Section \ref{sec:freq_fitting_uncertainty}.

\subsubsection{Finding detunings from Ramsey accumulated phase}

It is worthwhile to note that the phase difference $\Delta \phi$ accumulated during the (detuned) Ramsey sequence is \textit{not} equal to $f_\Delta \cdot \tau$ in practice (they are only equal in the limit where the pulse times for the rotations go to zero). 
To find the detuning $f_\Delta$ correctly then, we fit the measurement results to a unitary evolution given by

\begin{equation}
  \label{eq:unitary_ramsey_evolution}
  \ket{\psi_f} = \hat{U}_3\hat{U}_2\hat{U}_1\ket{\psi_0} = \hat{U}_3\hat{U}_2\hat{U}_1\ket{0}
  \hspace{1cm} \text{for} \hspace{1cm} \hat{U}_i = \exp(-i \hat{H}_i t).
\end{equation}

The two-level Hamiltonians for each step are

\begin{equation}
  \label{eq:hamiltonians_ramsey_evolution}
  \hat{H}_1 =
  \begin{pmatrix}0 & \frac{\Omega}{2} \\ \frac{\Omega}{2} & f_\Delta \end{pmatrix}
  \hspace{1cm}
  \hat{H}_2 =
  \begin{pmatrix}0 & 0 \\ 0 & f_\Delta \end{pmatrix}
  \hspace{1cm}
  \hat{H}_3 =
  \begin{pmatrix}0 & \frac{\Omega}{2}e^{i\phi} \\
    \frac{\Omega}{2}e^{-i\phi} & f_\Delta \end{pmatrix}.
\end{equation}

The actual calibration measurement involves only measuring the dark state probability at \textit{two points}, that is, with revival pulses set to $R(-\pi/2,\pi/2)$ and $R(-\pi/2,3\pi/2)$, as these two points are the most sensitive to detunings (steepest slopes in Supp. Fig.
\hyperref[suppfig:calibration_overview]{\ref*{suppfig:calibration_overview}c} at $f_\Delta=0$). 
We can call these measurement results $\{p_1,p_2\}$.
The detuning finder varies $f_\Delta$ and seeks to minimise the function

\begin{equation}
  \label{eq:cost_function_ramsey_calibration}
  \mathcal{C}(f_\Delta) = \abs{p_1 - \abs{\braket{0}{\psi_{f,\phi_1}}}^2}
  + \abs{p_2 - \abs{\braket{0}{\psi_{f,\phi_2}}}^2},
\end{equation}
where $\ket{\psi_{f,\phi'}}$ is the final state attained when the phase in $\hat{H}_3$ is $\phi=\phi'$. Of course, for our case, $\phi_1=\pi/2$ and $\phi_2=3\pi/2$. 

By waiting after this first pulse, any detuning between the laser frequency (which is effectively our guess for the transition frequency) and the true frequency of the transition will accumulate phase linearly with the length of time $\tau$ between the initial and final pulses, and therefore longer wait times are more sensitive to smaller detunings (so long as $\tau$ is less than the coherence time of the transition, which we discuss below).

\subsubsection{Frequency calibration uncertainty}
\label{sec:freq_fitting_uncertainty}

If we assume that we apply radiation at frequency $f_l$ for a transition with frequency $f_t$ such that the detuning $f_\Delta \equiv |f_l-f_t| \ll \Omega$,
for $\Omega$ the Rabi frequency of the transition, we can express the transition probability at the end of the Ramsey sequence (assuming purely Lorentzian laser noise) as

\begin{equation}
  \label{eq:transition_prob_ramsey_sequence}
  P(\phi) = \frac{1}{2}-\frac{1}{2}\cos(2\pi f_\Delta \cdot\tau+\phi) e^{-\tau/T_2^*}
\end{equation}
where $f_\Delta$ is the laser detuning, $\tau$ is the wait time between $\pi/2$-pulses, $\phi$ is the phase of the second $\pi/2$-pulse, and $T_2^*$ is the inhomogeneous dephasing time. 
Due to the finite number of shots taken experimentally, there will be some uncertainty associated with the measured transition probabilities at $\phi_1=\pi/2$ and $\phi_2=3\pi/2$, equal to $\sigma_P = \sqrt{p(1-p)/N}$, where $p$ is the probability and $N$ is the total number of shots in the measurement. 
We are interested in how the \textit{difference} in the two probabilities ($A$), which is what we use to find the frequency $f_\Delta$, impacts the uncertainty of this frequency measurement.

Since $A = P(3\pi/2)-P(\pi/2)$, the uncertainty related to measuring $A$ is $\sigma_A = \sqrt{2p(1-p)/N}$.
Meanwhile, at $\phi=\pi/2$ and $\phi=3\pi/2$,
\begin{equation}
  \label{eq:deriv_near_resonance_ramsey_sequence}
  \left|\frac{dA}{df_\Delta}\right| = 2\pi\tau e^{-\tau/T_2^*}.
\end{equation}

Then, since 
\begin{align}
    \sigma_{f_\Delta}^2 &= \sigma_A^2 / \left(\frac{dA}{df_\Delta}\right)^2 \\
    \sigma_{f_\Delta} &= \frac{\sqrt{2p(1-p)/N}}{2\pi\tau e^{-\tau/T_2^*}}.
\end{align}

The minimum uncertainty occurs where $\tau e^{-\tau/T_2^*}$ is maximised, at

\begin{align}
\label{eq:best_wait_time_ramsey_uncertainty}
  \frac{d}{dt}\tau e^{-\tau/T_2^*} &= e^{-\tau/T_2^*}\left( 1-\frac{t}{T_2^*} \right) = 0 \\
  \therefore t &= T_2^*.
\end{align}
As an example, using $N=250$, $\tau=\SI{100}{\micro\second}$, and a coherence time of $\SI{1}{\milli\second}$, and
assuming we are near resonance such that $p=0.5$, we would expect a detuning uncertainty of $\SI{78.6}{\hertz}$.

In order to get a realistic estimate for how well we can calibrate target transitions by only measuring $f_{int}$ and $f_{sens}$, we calculate the absolute deviations from the fit lines of each individual triplet measurement's value for $f_t$, for all 40 target transitions. 
Those deviations, for 520 total triplet measurements, are shown in Supp.~Fig.~\hyperref[suppfig:calibration_overview]{\ref*{suppfig:calibration_overview}c}. The inset shows the histogram of these measurement deviations, with a Gaussian fit which shows a standard deviation $\sigma_{fit}$ = \SI{126.3}{\hertz}. This value of the frequency miscalibration (listed as a full-width at half-max in table \ref{tab:noise_sources_distributions}) is used in all simulations with Monte Carlo sampling used in this paper.

\begin{figure} \centering
  \includegraphics[width=1.0\textwidth]{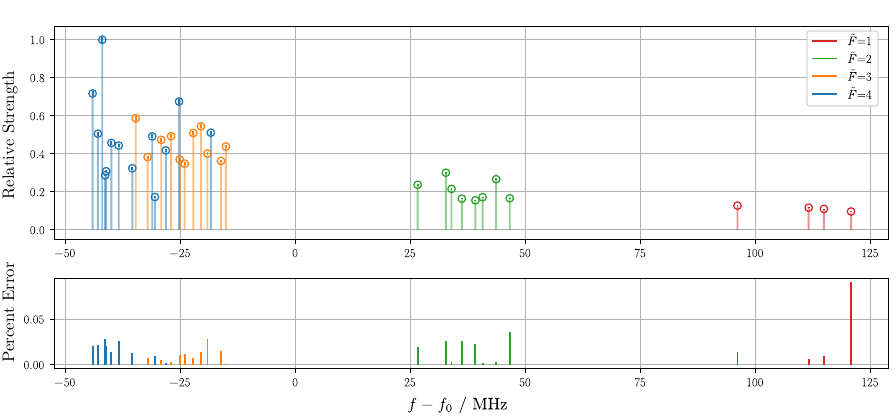}
  \caption[Calculated relative transition strengths, comparison with measured transitions.]
  {\textbf{Calculated relative transition strengths, comparison with measured transitions.} 
  Vertical lines in the top panel correspond to transitions (with frequencies on the $x$-axis calculated as in Section \ref{sec:transition_frequencies}) with heights denoting the relative strengths found from theory using five reference Rabi frequencies.
  Open circles denote the measured frequencies and strengths. 
  Lower panel shows the percent error between the measured and calculated transition strengths. 
  The mean absolute percent error for all transitions shown here is \SI{1.77}{\percent}.}
  \label{suppfig:transition_relative_strengths_errors}
\end{figure}

\subsection{Transition strengths}
\label{sec:transition_strengths}

In order to predict the relative strengths of the quadrupole transitions between the $S_{1/2}$ and $D_{5/2}$ manifolds, one cannot, as mentioned, assume pure hyperfine $\ket{F,m_F}$ eigenstates - especially in the $D_{5/2}$ manifold. 
We calculate the relative transition strengths for all transitions using the method outlined in Ref.~\cite{low25_contr_readout_trapp_ion_qudit} and compare those calculations to our measured results.

The measured results are shown in Supp.~Fig.~\ref{suppfig:transition_relative_strengths_errors}, with open circles denoting measured transition strengths, and bars denoting predicted strengths. 
The percent errors in each transition strength are shown in the lower panel. 
The mean absolute percent error in the transition strengths shown is \SI{1.77}{\percent} (referred to as $\delta\tau_c$ in the main text)).
For the data in Supp.~Fig.~\ref{suppfig:transition_relative_strengths_errors}, the reference transition strength used to predict a given target transition's strength is not always measured immediately preceding the measurement of the target transition's strength, and as a result, some slow drifts in eg. laser power fluctuations may also be contributing to the value of $\delta\tau_c$.
This percent error should be compared to the error expected when directly fitting Rabi oscillations, which is \SI{0.23}{\percent}. 
This difference is not surprising, as fluctuations in laser power and beam pointing likely contribute to changes in the measured Rabi frequencies, and thus apparent transition strengths. 
We further measure pulse-time drifts on the time scale of hours to be $\delta\tau_d =$ \SI{2.61}{\%}.

\section{NBOP Initialisation} 
\label{sec:2_s12_initialisation_choices}

\begin{figure} \centering
  \includegraphics[width=1.0\textwidth]{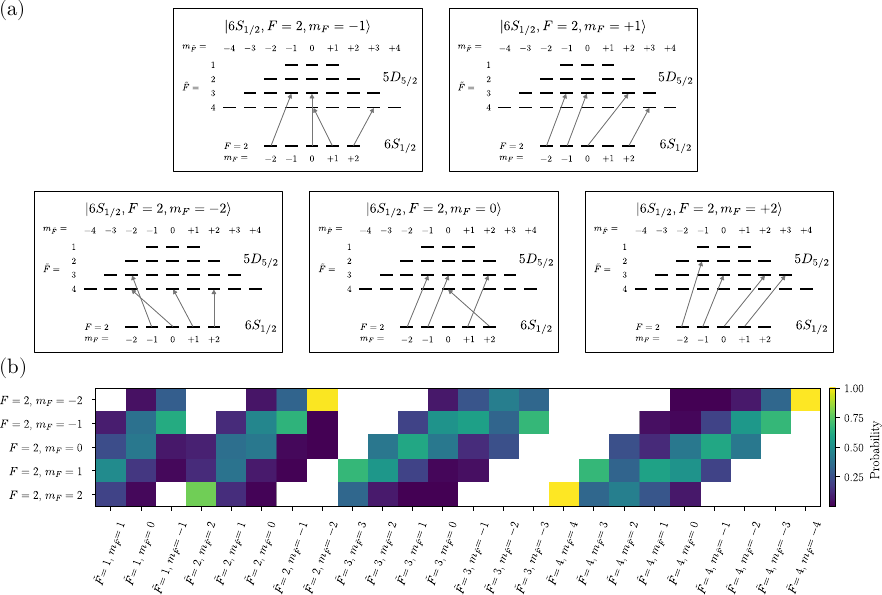}
  \caption[NBOP transition choices and repumping pathways probability map.]
  {\textbf{NBOP transition choices and repumping pathways probability map.} 
  \textbf{(a)} NBOP shelving transitions chosen to initialise $S_{1/2}, F=2$ states with high efficiency.
  \textbf{(b)} Re-pumping pathways probabilities found by calculating the transition matrix elements for $D_{5/2} \rightarrow$ $P_{3/2}$ states, assuming pure $\ket{F,m_F}$ states in $P_{3/2}$.}
  \label{suppfig:NBOP_state_choices_repumping}
\end{figure}

As discussed in Methods Sec. \ref{sec:A2_NBOP}, the transitions chosen for shelving in NBOP (which are shown in Supp.~Fig.~\hyperref[suppfig:NBOP_state_choices_repumping]{\ref*{suppfig:NBOP_state_choices_repumping}a}) have a big impact on the efficiency and resulting fidelity of the initialisation scheme. To derive the repumping pathways in Supp.~Fig.~\ref{suppfig:NBOP_state_choices_repumping}, we begin with the $D_{5/2}$ eigenstates for the magnetic field value at which experiments are run ($\sim$ \SI{4.21}{G}) and calculate the (normalised) probability that \SI{614}{\nano\meter} light pumps states to any of the 7 states in $P_{3/2}, F=3$. The \SI{614}{\nano\meter} beam is assumed to be equal parts $\sigma^+$ and $\sigma^-$ polarisations ($\pi$-polarisation is absent due to the beam being co-linear with the applied magnetic field axis).

Reduced matrix elements for decay from $P_{3/2}, F=3$ to $S_{1/2}, F=2$ can then straightforwardly be calculated, again assuming pure $\ket{F, m_F}$ states. Then, summing over the possible pathways from $D_{5/2}$ states back to $S_{1/2}$ and normalising gives the relative probabilities for repump paths shown in Supp.~Fig.~\hyperref[suppfig:NBOP_state_choices_repumping]{\ref*{suppfig:NBOP_state_choices_repumping}b}.

\section{25-level SPAM error calculations} 
\label{sec:3_SPAM_error_calculations}

Here we will consider the three contributions to SPAM infidelity in turn, and show how these contributions were taken together to model the infidelity as shown in Fig.~\hyperref[fig:d52_SPAM]{\ref*{fig:d52_SPAM}b}.

\subsection{Decay from meta-stable states}

The first error source is decay from the $D_{5/2}$ levels back to the ground states. 
Each fluorescence check integrates over PMT counts for \SI{5}{\milli\second}, which means that, after initialisation, a state in $D_{5/2}$ is waiting to be de-shelved back during sequential readout for up to \SI{125}{\milli\second} (depending on its position in the de-shelving order). 
From the lifetime of the $D_{5/2}$ levels of \SI{30.1(4)}{\second}~\cite{zhang20_branc_fract}, we can directly calculate the expected spontaneous decay probability for each state as $P_d = e^{-t/\tau}$ where $t = n \cdot \SI{5}{\milli\second}$. 
This yields a maximum fidelity of $1-P_d = \SI{99.983}{\percent}$ when $n=1$, and \SI{99.60}{\percent} when $n=24$, leading to an average infidelity of $1.92 \times 10^{-3}$ from this error source.

\subsection{Off-resonant driving}

\begin{figure*}
\centering
\includegraphics[width=0.9\textwidth]{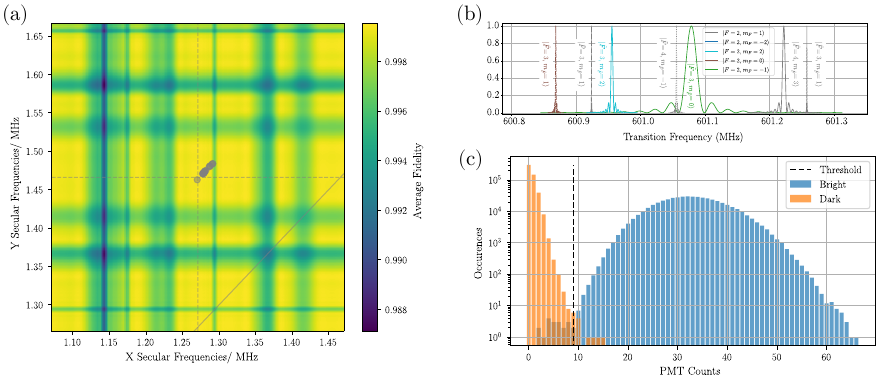}
  \caption[Contributors to 25-level SPAM error.]{\textbf{Contributors to 25-level SPAM error.} \textbf{(a)} Calculated SPAM fidelities as a function of $x-$ and $y-$secular frequencies of the trap. Calculations made with $\eta = 0.014$, $\overline{n}=140$ (previously measured on this system~\cite{low23_contr_reado_high_dimens_trapped}), axial-secular frequency of \SI{215}{\kilo\hertz}, and magnetic field value of \SI{4.209}{G}. Gray dots in the center of the plot illustrate experimental measurements of the (correlated) secular frequency drifts over several hours. \textbf{(b)} Simulated transition Rabi fringes for a small portion of the full $S_{1/2} \leftrightarrow D_{5/2}$ transition map, intended to illustrate the (possible) contributions of off-resonantly driven transitions, largely through motional sideband transitions (dashed lines). \textbf{(c)} Readout histogram of PMT counts collected during a \SI{5}{\milli\second} exposure with no ion present (dark) and with an ion present (bright). Both fluorescent readout lasers (\SI{493}{\nano\meter} and \SI{650}{\nano\meter} are on for both measurements. The ideal threshold of 9 is denoted by the dashed black line.}  \label{suppfig:d52_SPAM_errors_supplementary}
\end{figure*}

Another source for error stems from off-resonant excitation of unwanted $S_{1/2}\leftrightarrow D_{5/2}$ transitions. 
We have calculated that the lowest spacing between adjacent transitions (including motional sideband transitions) can be as low as $\sim \SI{50}{\kilo\hertz}$ (see Supp.~Fig.~\hyperref[suppfig:d52_SPAM_errors_supplementary]{\ref*{suppfig:d52_SPAM_errors_supplementary}b}), and
the off-diagonal elements in Fig.~\ref{fig:d52_SPAM} show evidence of off-resonant driving (for instance, between the states $\ket{5}$ and $\ket{21}$). 
In order to estimate the contribution from this error source we perform a calculation as follows:

\begin{enumerate}
    \item Determine the frequencies and transition strengths of all $S_{1/2} \leftrightarrow D_{5/2}$ transition pathways.
    
    \item For each \textit{shelving} pulse intended to move population to a desired $D_{5/2}$ initial state in the SPAM measurement, calculate the probability for off-resonantly driving population to an undesired $D_{5/2}$ state using the well-known Rabi fringe expression

    \begin{equation}
      \label{eq:rabi_lineshape}
      P_i(f,t) = \frac{\Omega_i^2}{\Omega_i^2+(f-f_i)^2} \sin^2
      \left(\sqrt{\Omega^2+(f-f_i)^2} \pi t \right),
    \end{equation}
    where $\Omega_i$ is the Rabi frequency and $f_i$ is the resonant transition frequency for that particular transition $i$, and $t$ is the time for which the pulse is applied.
    Using these probabilities, re-normalise the probability for a successful shelving pulse for that particular state.
    For this step in the calculation, only transitions that couple to the $S_{1/2}$ state from which the desired $D_{5/2}$ state is shelved are relevant (i.e. sharing the same column in Supp.~Fig.~\ref{suppfig:transition_frequencies_sensitivities}), as there will be very little population in all other $S_{1/2}$ states when the shelving pulse is applied.

    Put explicitly, we find the set of all transitions that couple resonantly to the initialised $S_{1/2}$ state, call this $\{s\}$, each with frequencies $f_s$ (as well as axial and radial sideband frequencies).
    We treat all probabilities of off-resonant driving to be small such that they can be treated as each contributing linearly to the decrease in the shelving fidelity $\mathcal{F}_s$.
    The total fidelity then, when attempting to shelve population for $\ket{i}$ using a transition of frequency $f_i$ and pulse time $t$, can be found as

    \begin{equation}
        \mathcal{E}_{\{s\}} = \sum\limits_{\{s\}} P_i(f_s,t).
    \end{equation}

    \item When \textit{de-shelving} states from $D_{5/2}$, we now need to take into account the measurements order for each state $\ket{i}$. Since the measurement procedure attempts to de-shelve population in the order depicted in Fig.~\hyperref[fig:d52_SPAM]{\ref*{fig:d52_SPAM}d}, each such de-shelving pulse has some probability of off-resonantly driving population back to $S_{1/2}$.

    For a given initialised state $\ket{i}$ then, the set of transition pulses sent to de-shelve population from $\ket{1}$ up to $\ket{i-1}$ we can call $\{d\}$. 
    Each of these de-shelving pulses have their own transition times $t_d$ for which the pulses are played, and furthermore, off-resonant de-shelving of our initialised population to \textit{any} $S_{1/2},F=2$ level will yield an error in our SPAM result, so we need to account for the probability that these preceding de-shelving pulses pull down the initialised $D_{5/2}$ state via any of its available transitions to $S_{1/2}$ states (shown as rows in Supp.~Fig.~\ref{suppfig:transition_frequencies_sensitivities}).
    We therefore sum the probability over all preceding de-shelving pulses in $\{d\}$, but also over the set of transitions from the initialised $D_{5/2}$ state back to $S_{1/2}$, which we can denote as $\{q\}$.

    The total error, taking into account de-shelving only, can be expressed as

    \begin{equation}
        \mathcal{E}_{\{d\}} = \sum\limits_{\{d\}} \sum\limits_{\{q\}} P_q(f_s,t).
    \end{equation}
    
    we perform the same procedure calculating probabilities that other states' populations in $D_{5/2}$ will be off-resonantly pumped back to $S_{1/2}$.

    \item Taking into account these two contributions (from the \enquote{preparation} and \enquote{measurement} pulses separately then summing the results), we calculate the expected error due to off-resonant driving for each $\ket{i}$: $\mathcal{E}_{\ket{i}} = \mathcal{E}_{\{s\}} + \mathcal{E}_{\{d\}}$, and we can then average this error over all the states participating in the SPAM measurement.
    Each state will have higher or lower re-normalised probabilities to be off-resonantly driven, and these also will depend on the secular frequencies of the trap, as sideband transitions can shift into and out-of resonance with carrier transitions of other states.
    
\end{enumerate}

In Supp.~Fig.~\hyperref[suppfig:d52_SPAM_errors_supplementary]{\ref*{suppfig:d52_SPAM_errors_supplementary}a}, we have calculated the expected 25-level SPAM fidelity as a function of the two secular frequencies in the trap. 
Overlayed grey dots indicate a series of measurements of the trap secular frequencies over a 5-hour period, which indicate the degree to which these frequencies are stable of time scales similar to those required to take the SPAM measurements in the main text.
This degree of fluctuations in the secular frequencies, as well as their form (which is indicative of power fluctuations in the delivered trap-RF signal through our $\sim \SI{20.7}{\mega\hertz}$ resonator) are also fed into the model for calculating SPAM infidelity from this error source.
We find a contribution of $2.03 \times 10^{-3}$ from off-resonant driving in the SPAM infidelities for $d=25$.

\subsection{Bright/dark state discrimination}

Finally, imperfect bright/dark state discrimination contributes to our infidelity. 
This can arise from dark counts in the photo-multiplier tube pushing the photons counted per readout above the threshold used to distinguish between a bright and dark outcome, or from a failure to collect enough photons from a bright state to pass the threshold. 
The histogram of readout outcomes for this procedure features two peaks with some overlap (see Supp.~Fig.~\hyperref[suppfig:d52_SPAM_errors_supplementary]{\ref*{suppfig:d52_SPAM_errors_supplementary}c}), indicating that some mis-attribution of readout outcomes is occurring.
We measure a bright/dark discrimination error (combining false positive and false negative errors) of $5.6 \times 10^{-5}$ at a threshold of 9 PMT counts per \SI{5}{\milli\second} exposure.

We conclude this discussion of the error sources in SPAM by noting that, though the error estimate fits the experimental data well, it is still a slight underestimate of the total error measured here (approximately $3 \times 10^{-4}$ remains unaccounted for). 
We can point to at least one additional noise source not discussed above, which is ion cooling. 
As mentioned previously, all experiments are done using Doppler cooling only, and we have found that insufficient cooling can lead to SPAM errors in which population initialised in the state $\ket{n}$ is measured to be in $\ket{n+1}$ with non-negligible probability. 
This happens because an insufficiently cool ion may not fluoresce as efficiently, leading to sub-threshold outcomes for the number of collected photons during the readout of state $\ket{n}$.
(Indeed, there are some indications of this in Fig.~\hyperref[fig:d52_SPAM]{\ref*{fig:d52_SPAM}c}.)

\section{Noise measurements and modelling} 
\label{sec:4_system_noise_studies}

\begin{figure} \centering
\includegraphics[width=1.0\textwidth]{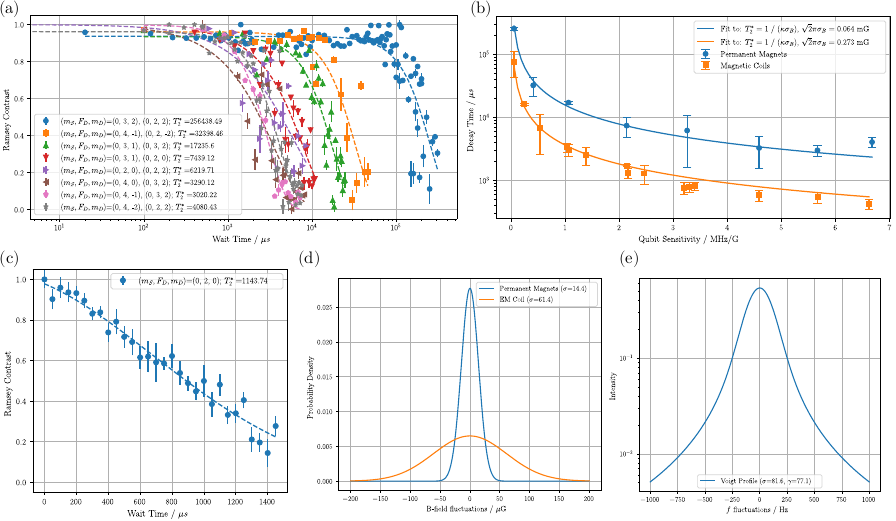}
\caption[Magnetic field and laser noise characterisation.]{\textbf{Magnetic field and laser noise characterisation.} \textbf{(a)} Decays in Ramsey contrast measurements as a function of wait times in superpositions for pairs of states in $D_{5/2}$. \textbf{(b)} Two-level Ramsey contrast decay times plotted against the relative magnetic field sensitivities of the states involved. We extract values for the magnetic field fluctuations based on these results. \textbf{(c)} Contrast decay for the magnetic field insensitive transition ($\ket{S_{1/2}, F = 2, m_F = 0}$ to $\ket{D_{5/2}, \tilde{F} = 2, m_{\tilde{F}} = 0}$) which fits to laser noise with a Voigt profile. Example probability density functions for magnetic field noise \textbf{(d)} and laser noise \textbf{}(e) estimated from the measured data in \textbf{(a-c)} show the distributions sampled for simulations of system evolution using the Monte-Carlo sampling method described in Methods \ref{sec:A5_noise_simulations}.}
  \label{suppfig:noise_plots}
\end{figure}

\subsection{Magnetic field noise}

To characterize random magnetic field noise, we perform Ramsey experiments on pairs of states both in the $D_{5/2}$ manifold so that the time dependence of the phase decoherence is free from laser frequency noise (see Sec. \ref{sec:laser_free_decoherence}).
We observed that the phase coherence decays with a Gaussian profile, as shown in Supp.~Fig.~\hyperref[suppfig:noise_plots]{\ref*{suppfig:noise_plots}a}, signifying that the random distribution of the magnetic field is Gaussian, which we have also verified with an independent direct measurement of the magnetic field.
From our experiments, we measure a Gaussian standard deviation of \SI{14.4}{\micro G} for the random magnetic field noise (blue curve in Supp.~Fig.~\hyperref[suppfig:noise_plots]{\ref*{suppfig:noise_plots}b}) when using permanent magnets to generate the field. We also performed this measurements when using current carrying copper coils to generate the magnetic field, where we found \SI{61.4}{\micro G} of fluctuations (over a four-fold increase as compared to magnets).
Supp.~Fig.~\hyperref[suppfig:noise_plots]{\ref*{suppfig:noise_plots}d} shows the difference between the estimated probability density functions for each magnetic field source.

\begin{figure} \centering
\includegraphics[width=1.0\textwidth]{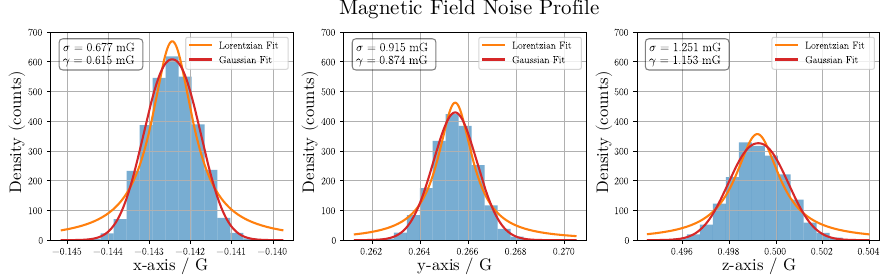}
\caption[Magnetic field noise, outside vacuum chamber.]{\textbf{Magnetic field
    noise, outside vacuum chamber.} Magnetic field measurements as taken by a
  magnetometer outside the vacuum chamber, with the $y$-axis aligned with the
  quantisation axis of the system. We find that the noise profile in all
  directions is well described by a standard normal distribution.}
  \label{suppfig:no_shield_field_noise_profile}
\end{figure}

We independently verify the profile of the magnetic field noise in the lab by measuring, outside the trap, fluctuations in the field using a commercially available magnetometer (MEMSIC MMC5983MA). The results are best described as Gaussian, as expected (Supp.~Fig.~\ref{suppfig:no_shield_field_noise_profile}).

\subsubsection{Laser frequency noise decoherence cancellation}
\label{sec:laser_free_decoherence}

In this section, we evaluate how decoherence due to laser frequency noise can be canceled by doing an even number of Givens rotations, given that the computational states are both lower or both higher than the bus states in energy levels.
To evaluate how coherence times are affected between two states, let them be $\ket{0}$ and $\ket{1}$, that are connected via a bus state, $\ket{B}$, we consider a usual Ramsey sequence consisting of four ideal two-level transition unitaries as follows:
\begin{equation}
    \begin{aligned}
        \ket{\psi} &= \hat{U}_{2,1} \hat{U}_{2,2} \hat{U}_{1,2} \hat{U}_{1,1} \ket{0} \\
        \hat{U}_{1,1} &= \sqrt{\frac{1}{2}} \left( \ket{0} \bra{0} + \ket{B} \bra{B} + \ket{B} \bra{0} - \ket{0} \bra{B} \right) + \ket{1} \bra{1} \\
        \hat{U}_{1,2} &= \ket{1} \bra{B} - \ket{B} \bra{1} + \ket{0} \bra{0} \\
        \hat{U}_{2,2} &= -\ket{1} \bra{B} + \ket{B} \bra{1} + \ket{0} \bra{0} \\
        \hat{U}_{2,1} &= \sqrt{\frac{1}{2}} \left( \ket{0} \bra{0} + \ket{B} \bra{B} - \ket{B} \bra{0} + \ket{0} \bra{B} \right) + \ket{1} \bra{1}.
    \end{aligned}
\end{equation}
To investigate the effect of decoherence, we incorporate time-varying frequency detunings of the transitions into the model.
We keep the assumption that the pulse time is short, $\Delta \omega_n \delta t \ll 1$, such that we can assume the effective Hamiltonian to be time-independent when the laser is turned on.
The effect of the time-varying frequency detunings is then to shift the phase of the transition at some time $t$ by $\int_0^t \Delta \omega_n \left( t' \right) dt'$, such that the effective Hamiltonian can be approximated to be
\begin{equation}
    \hat{H}_I \approx \frac{\Omega_m}{2} \left( i e^{i \int_0^t \Delta \omega_m \left( t' \right) dt' + \phi_m} \ket{B} \bra{m} - i e^{-i \int_0^t \Delta \omega_m \left( t' \right) dt' - \phi_m} \ket{m} \bra{B} \right)
\end{equation}
and the unitary generated by the Hamiltonian is
\begin{multline}
    \hat{U} = \ket{n} \bra{n} + \cos \left( \frac{\Omega_m}{2} \delta t \right) \left( \ket{0} \bra{0} + \ket{m} \bra{m} \right) \\
    + \sin \left( \frac{\Omega_m}{2} \delta t \right) \left( - e^{i \int_0^t \Delta \omega_m \left( t' \right) dt' + \phi_m} \ket{B} \bra{m} + e^{-i \int_0^t \Delta \omega_m \left( t' \right) dt' - \phi_m} \ket{m} \bra{B} \right).
\end{multline}
Evaluating $\bra{0} \psi \rangle$ gives
\begin{equation}
    \begin{aligned}
        \bra{0} \hat{U}_{2,1} \hat{U}_{2,2} \hat{U}_{1,2} \hat{U}_{1,1} \ket{0} &= \frac{1}{2} \left( 1 + e^{i \left( \int_{t_2}^{t_3} \Delta \omega_1 \left( t' \right) dt' - \int_{t_1}^{t_4} \Delta \omega_0 \left( t' \right) dt' \right)} \right) \\
    \end{aligned}
    \label{eq:Bussed_Ramsey_general_error}
\end{equation}
where each of the four pulses end at times $t_1, t_2, t_3, t_4$.
Suppose that the sources of frequency detunings only come from laser frequency noise and magnetic field noise, and the computational states are both higher or both lower than the bus state  we can decompose the frequency detunings to
\begin{equation}
    \Delta \omega_n = \Delta \omega_L \left( t' \right) + \kappa_n \Delta B \left( t' \right).
    \label{eq:Frequency_detuning_laser_Bsensitivity_breakdown}
\end{equation}
From eqs. \ref{eq:Bussed_Ramsey_general_error} and \ref{eq:Frequency_detuning_laser_Bsensitivity_breakdown}, we have
\begin{equation}
    \begin{aligned}
        \bra{0} \hat{U}_{2,1} \hat{U}_{2,2} \hat{U}_{1,2} \hat{U}_{1,1} \ket{0} &= \frac{1}{2} \left( 1 + e^{i \left( A \left( \kappa_0 \right) + \int_{t_2}^{t_3} \left( \kappa_1 - \kappa_0 \right) \Delta B \left( t' \right) dt' \right)} \right),
        \label{eq:Ramsey_measurement_outcome_laser_noise_canceled}
    \end{aligned}
\end{equation}
where
\begin{equation}
    A \left( \kappa_0 \right) = -\int_{t_1}^{t_2} \Delta \omega_L \left( t' \right) dt' -\int_{t_3}^{t_4} \Delta \omega_L \left( t' \right) dt' - \int_{t_1}^{t_2} \kappa_0 \Delta B \left( t' \right) dt' - \int_{t_3}^{t_4} \kappa_0 \Delta B \left( t' \right) dt'
\end{equation}
is the component independent of the wait time between $\hat{U}_1$ and $\hat{U}_2$. 
From eq. \ref{eq:Ramsey_measurement_outcome_laser_noise_canceled}, it can be seen that only the magnetic field noise contribute to decay in coherence between the time $t_2$ and $t_3$, and laser frequency noise is a non-factor.

\subsection{Laser noise}

To characterize the laser frequency noise in our system, we performed Ramsey experiments on a magnetically insensitive transition ($ < \SI{1}{\kilo\hertz / G}$), which is the $\ket{S_{1/2}, F = 2, m_F = 0}$ to $\ket{D_{5/2}, \tilde{F} = 2, m_{\tilde{F}} = 0}$ transition, again extracting the contrast of the oscillation as a function of the wait time in superposition (Supp.~Fig.~\hyperref[suppfig:noise_plots]{\ref*{suppfig:noise_plots}c}).
We observed that the coherence does not decay purely exponentially with the wait time for the second Ramsey pulse, but with the characteristic function of a Voigt distribution.
This implies that the randomness of the laser frequency noise has a Gaussian component in addition to the expected Lorentzian distribution.
Thus, we fit the Ramsey coherence decay data with the function
\begin{equation}
    y = A e^{-t^2/T^2_{L,G} - t/T_{L,L}},
\end{equation}
where $A$ is the initial contrast at zero wait time between the Ramsey pulses, $T_{L,G}$ and $T_{L,L}$ are the characteristic phase coherence decay times of the Gaussian and Lorentzian components of the laser frequency noise respectively.
From the fit we obtain the parameters characterizing laser frequency noise of $T_{L,G} = \SI{2065.5}{\micro \second}$ and $T_{L,L} = \SI{1950}{\micro \second}$.
These values are then used to extract noise estimates for the Gaussian component (\SI{81.6}{\hertz}) and Lorentzian component (\SI{77.1}{\hertz}), as shown in Supp.~Fig.~\hyperref[suppfig:noise_plots]{\ref*{suppfig:noise_plots}e}.

\subsection{A/C line signal}

We measure the A/C line signal by measuring the frequencies of one magnetic field sensitive transition, and the least magnetic field sensitive transition of all the available \SI{1762}{\nano\meter} transitions. 
From those two frequencies we are able to extract a magnetic field value based on theory. 
This magnetic field value serves as an initial reference point. 
This initial reference point has both frequency measurements triggered on the rising edge of the A/C line signal. We then repeat this measurement for a range of wait times after the A/C line signal rising edge, going from 0 to \SI{16.6}{\milli\second} to sample the full \SI{60}{\hertz} period of the A/C line signal.

\begin{figure} \centering
\includegraphics[width=1.0\textwidth]{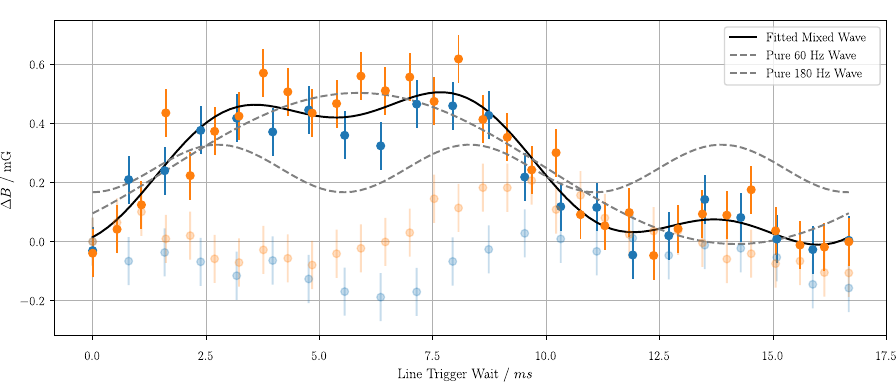}
\caption[A/C line signal measurement.]{\textbf{A/C line signal measurement.} Measured change in magnetic field as a function of the line A/C mains line signal phase. The two main components to this signal are at \SI{60}{\hertz} (amplitude \SI{0.128}{\milli G}, phase \SI{-0.636}{\radian}) and \SI{180}{\hertz} (amplitude \SI{0.04}{\milli G}, phase \SI{-1.55}{\radian}). Blue and orange data points are phase scans measured 2 months apart. Faded data points are interleaved reference measurements of the magnetic field at \SI{0}{\milli\second} wait time for the line signal, which captures and disentangles the slow drift in magnetic field throughout this line signal measurement.}
  \label{suppfig:ac_line_signal}
\end{figure}

We find two main components to the line signal, with frequencies of \SI{60}{\hertz} and the third harmonic at \SI{180}{\hertz}, as shown in Supp.~Fig.~\ref{suppfig:ac_line_signal}. 
Each component is plotted using dashed grey lines, with the total fit in black. 
The \SI{60}{\hertz} signal has an amplitude of \SI{0.128}{\milli G} and phase of \SI{-0.636}{\radian}, and the \SI{180}{\hertz} component is fit to an amplitude of \SI{0.04}{\milli G}, with a phase of \SI{-1.55}{\radian}. 
The blue and orange data points are two separate full scans taken two months apart (to confirm the stability of the signal). 
Each measurement of the magnetic field at a certain wait time is interleaved with a reference measurement at 0 wait time (faded blue and orange data points) in order to disentangle slow magnetic field drifts from the influence of the line signal.

\section{Towards high-fidelity qudits}
\label{sec:5_high_fidelity_qudits}

In general, the results of this work have shown that as the number of states involved in a computation increases (ie. as the dimension of qudit or polyqubit increases) the fidelity of the results tends to fall off. 
This naturally leads to the question of what levels of noise in the system are required in order to nonetheless be able to implement high-fidelity operations. 

In order to gain ground with this question, we must choose a concrete target, for which we set a goal of $10^{-4}$ infidelity. We wish to answer the question: what levels of magnetic field noise, laser noise, frequency/power (mis-)calibration, and A/C line signal are acceptable for a combined $10^{-4}$ contrast loss as a function of the dimension $d$ of the qudit in question?

\subsection{High-Fidelity SPAM}
As mentioned in the main text, as well as in \ref{sec:3_SPAM_error_calculations}, there are three main contributions to SPAM error: a) decay from $D_{5/2}$ to the ground $S_{1/2}$ levels, b) bright/dark discrimination error, and c) off-resonant driving between encoded states. 
We will discuss what hardware requirements must be met to ensure that each of these error sources contribute at most one-third of the total $10^{-4}$ error rate.

\begin{enumerate}
    \item \textbf{Spontaneous decay:} To limit this error the fluorescence readout time in the experiment must be reduced. By directly calculating the probability for $D_{5/2}$ state decay during multiple sequential readouts, for a given readout time $\tau_{RO}$, given a lifetime of \SI{30.1(4)}{\second}~\cite{zhang20_branc_fract}, we calculate a contribution to our SPAM measurement of $\epsilon_{decay} = 1.92\times 10^{-3}$ with our $\tau_{RO}=\SI{5}{\milli\second}$. 
    We find that a readout time of \SI{80}{\micro\second} is short enough to keep error from this source to $3.2\times 10^{-5}$. 
    This is much shorter than the \SI{5}{\milli\second} fluorescence time used in this work, but not unreasonable for systems with better photon collection efficiencies~\cite{an22_high_fidel_state_prepar_measur, sotirova24_high_fidel_heral_quant_state_prepar_measur}.

    \item \textbf{Bright/dark state discrimination:} This shorter photon collection time for fluorescent readout must be paired with a lower bright/dark discrimination error rate. 
    The current bright/dark discrimination error in our system is $5\times 10^{-5}$, which yields an overall $\epsilon_{RO} = 6.25\times 10^{-4}$ for $d=25$. 
    Reducing the bright/dark discrimination to $2.5\times 10^{-6}$ (again, by increasing photon collection efficiency through optimised trap geometry) would bring down infidelity from this error source to $3.1 \times 10^{-5}$. This can be achieved by collecting ~40 photons when bright, which for a readout time of \SI{80}{\micro\second}, and a scattering rate for \SI{493}{\nano\meter} of \SI{2}{\mega\hertz} (\SI{10}{\%} of the linewidth of the $S_{1/2}\leftrightarrow P_{1/2}$ transition), requires a combined collection and detector efficiency of $\sim \SI{25}{\%}$.
    Assuming a detection efficiency $\eta_{detect} \sim 0.9$, this implies a numerical aperture of NA = 0.89.

    This high a numerical aperture may be impractical in many systems, but if we relax the requirement somewhat to target $10^{-3}$ error rate, we can increase the photon collection time to \SI{800}{\micro\second} (twice that of Ref.~\cite{sotirova24_high_fidel_heral_quant_state_prepar_measur}) while keeping error from spontaneous decay to $3\times 10^{-4}$, thus allowing a combined collection and detection efficiency of \SI{2.5}{\%}.
    This then implies that, even with a much more modest $\eta_{detect} \sim 0.5$, a NA = 0.43 is sufficient for the \SI{5}{\%} photon collection efficiency required.

    \item \textbf{Off-resonant driving:} The error source fundamentally stems from the density of states in the $D_{5/2}$ manifold, all of which are within around \SI{150}{\mega\hertz} of one another. 
    However, many other parameters also influence the degree of off-resonant driving (mostly through motional sideband transitions) in the system - such as the Lamb-Dicke parameter $\eta$, the average phonon number $\overline{n}$, as well as the magnetic field which influences the spread in carrier transition frequencies. 
    As a result, there are likely many different sets of system parameters that could bring off-resonant driving to below a $10^{-4}$ contribution to SPAM infidelity. 
    For instance, simply bringing down the average phonon number to $\overline{n}=5$, with the same Lamb-Dicke parameter of $\eta = 0.014$, we find that at a magnetic field value of $B = \SI{10.67}{G}$ the error introduced by off-resonant driving falls below $1.5\times 10^{-4}$. 
    Further optimisations of the trap secular frequencies and more effective ion cooling (to bring $\overline{n}$ down further, through, for example, electromagnetically induced transparency (EIT) cooling~\cite{huang24_elect_induc_trans_coolin_high}) could lead to yet lower off-resonant driving effects.
\end{enumerate}

\subsection{High-contrast qudit Ramsey control}

\begin{figure} \centering
\includegraphics[width=1.0\textwidth]{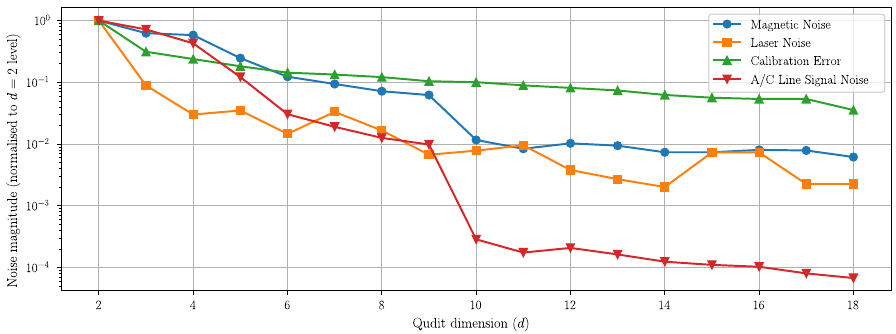}
\caption[Noise levels necessary for $10^{-4}$ level errors.]{\textbf{Noise levels necessary for $10^{-4}$ level errors.} Turning on each noise source alone, and for each dimension from $d=2$ to 18, we estimate how the noise sources must scale as a function of qudit dimension to maintain less than $10^{-4}$ contrast loss. Values are normalised to those found for $d=2$. Calibration error is frequency error only, we assume no pulse timing error for these simulations.}
  \label{suppfig:qudit_Ramsey_99p99_noises}
\end{figure}

The noise model outlined in Methods \ref{sec:A5_noise_simulations} to capture the dynamics of the qudit Ramsey measurements allows one not only to reproduce the measured results with good agreement (as shown in Fig.~\hyperref[fig:qudit_Ramsey_phase_scans_contrasts]{\ref*{fig:qudit_Ramsey_phase_scans_contrasts}e}), but it also allows for the study of individual noise sources. Using our model, we are able to simulate the behaviour of isolated noise sources and their effect on our system. 

Though the qudit Ramsey contrast which we measure does not map directly onto single qudit gate fidelities, we nonetheless believe it to be a good heuristic measure for estimating future gate fidelities. Here, we discuss the levels of noise reduction needed in our system to maintain contrast losses below the $10^{4}$ level, and we will focus on the $d=16$ case in particular.

The noise levels calculated in Fig.~\ref{suppfig:qudit_Ramsey_99p99_noises} are referenced to the noise values required for $10^{-4}$ contrast loss on a qubit encoding, and each assumes that only \textit{one} noise source is present in the system at a time. The qubit ($d=2$) noise values found are (1) \SI{58.6}{\micro G} for $\sigma_B$ Gaussian standard deviation for magnetic field fluctuations, (2) \SI{56.1}{\hertz} Voigt profile half-width at half-max laser frequency/phase noise, (3) \SI{227}{\hertz} frequency calibration error, and (4) \SI{9.3}{\milli G} A/C line signal amplitude. All noise sources need to be reduced by at least an order of magnitude to maintain contrast up to $d=18$ (we stopped these calculations at $d=18$ as the feedback optimisation over the noisy Monte-Carlo sampled outcomes of the simulation became computationally expensive). The required values at $d=16$ were found to be (1) \SI{0.47}{\micro G} for $\sigma_B$ Gaussian standard deviation for magnetic field fluctuations, (2) \SI{0.41}{\hertz} Voigt profile half-width at half-max for laser frequency/phase noise, (3) \SI{12.2}{\hertz} frequency calibration error, and (4) \SI{0.96}{\micro G} A/C line signal amplitude. 

\begin{figure} \centering
\includegraphics[width=1.0\textwidth]{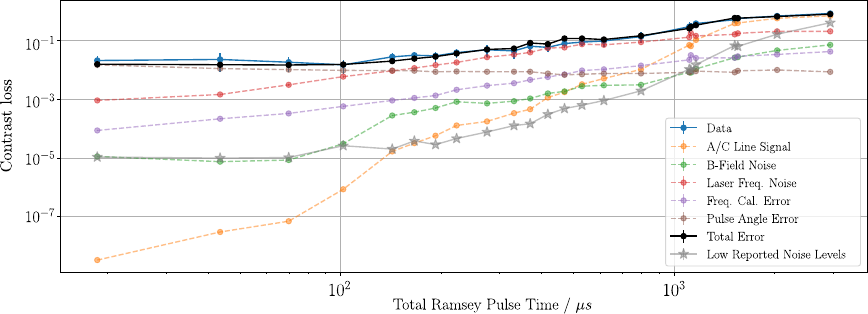}
\caption[Qudit Ramsey contrast measurements versus pulse times.]{\textbf{Qudit Ramsey contrast measurements versus pulse times.} Contrasts of the $\ket{0}$ state populations as a function of qudit dimension $d$, along with the prediction based on known noise sources affecting the ion. 
The sub-plots represent the same data on a linear scale (top) and log scale (bottom).
The log scale plot includes simulated qudit Ramsey results with lower noise values reported in literature.
See Methods \ref{sec:A3_transition_state_choice} for physical state choices and transitions used for each dimension of the qudit Ramsey measurements.
Error bars denote $1\sigma$ confidence using the Wilson interval~\cite{wilson27_probab_infer_law_succes_statis_infer}.}
  \label{suppfig:qudit_Ramsey_contrasts_vs_pulsetimes}
\end{figure}

It is also worth noting here that contrast loss could further be mitigated by reducing the overall run times of the qudit Ramsey pulse sequences, as Supp.~Fig.~\ref{suppfig:qudit_Ramsey_contrasts_vs_pulsetimes} shows. We can clearly see here that the contrast loss with higher dimension qudit Ramsey measurements correlates with the longer pulse sequence times needed to the lengthier multi-pulse rotations required for $d>17$.

\subsection{Qudit control with known noise levels}

Some of the noise values quoted in the above section do not yet have matching literature values and so, in order to provide the reader with intuition on what the attainable level of control could be using currently reported technology, we have compiled a list of noise levels in Table \ref{tab:known_best_noise_sources}. 
We use these reported values, along with some assumptions on the frequency and pulse time calibration results achievable in this future low-noise system (\SI{5}{\hertz} frequency miscalibration, and \SI{0.1}{\%} pulse timing miscalibration and pulse timing drift) and find that our system simulation for the $d=16$ qudit Ramsey result yields a contrast loss of $1.0(2)\times 10^{-3}$.

We also use the same noise values in the simulation for the Bernstein-Vazirani algorithm and find a mean failure probability of $4(3)\times 10^{-3}$.

\begin{table}
    \vspace{2em}
    \centering
    \begin{tabular}{|c|c|c|}
    \hline
        Noise Type & Value & Reference \\
        \hline
         Magnetic Field & $\SI{0.0382}{\micro G}$ &~\cite{ruster16_long_lived_zeeman_trapp_ion_qubit} \\
         Laser Frequency & \SI{0.5}{\hertz} &~\cite{young99_visib_laser_with_subher_linew, alnis08_subher_linew_diode_laser_by} \\
         A/C Line Signal (\SI{60}{\hertz} only) & $\SI{70}{\micro G}$ &~\cite{ruster16_long_lived_zeeman_trapp_ion_qubit} \\
         \hline
    \end{tabular}
    \caption{\textbf{Noise sources, profiles, and distribution widths.} Above we show the various contributions to detuning and pulse-time errors in the Monte Carlo simulations of qudit Ramsey-type experiments and the Bernstein-Vazirani algorithm. Further discussion on the measurements leading to these profiles and widths can be found in Supp. Section \ref{sec:4_system_noise_studies}. The Voigt profile width reported here is calculated from the decay of a magnetically insensitive transition.}
    \label{tab:known_best_noise_sources}
\end{table}

\begin{figure} \centering
\includegraphics[width=1.0\textwidth]{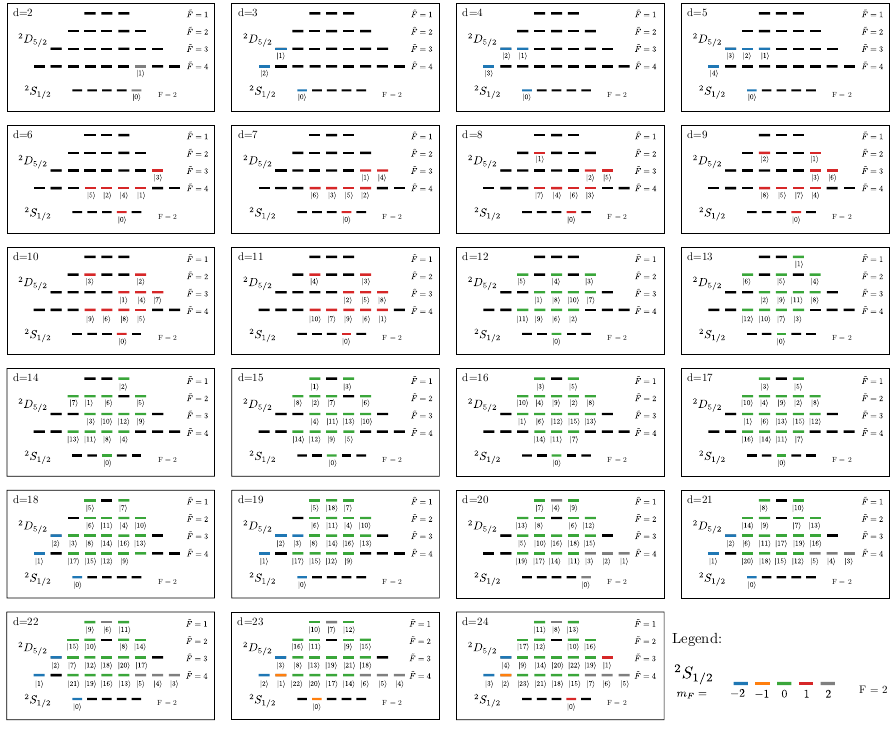}
\caption[Full encoding schemes for all dimensions of qudit Ramsey measurements.]{\textbf{Full encoding schemes for all dimensions of qudit Ramsey measurements.} State encoding schemes for all qudit Ramsey experiments, from $d=2$ to 24. The colours for each encoded state indicate the transition used to reach that encoded state from $S_{1/2}$, see the legend in lower right of the figure (black lines are un-encoded states). For $d \leq 17$ all states share a single $S_{1/2}$ level (star-type topology). For $d > 18$, some multi-transition pulses are required to reach a larger number of states, necessitating transitions to other levels in $S_{1/2}$.}
  \label{suppfig:qudit_Ramsey_all_encoding_schemes}
\end{figure}

\begin{figure}
  \centering
  \mbox{ \Qcircuit @C=1em @R=2em {
\lstick{\ket{00}} & \qw & \sgate{R_y(-\pi/2)}{3} & \sgate{R_y(\theta_1)}{1} & \sgate{R_y(-4\pi/3)}{2} & \sgate{R_y(\pi+\theta_2)}{1} & \sgate{R_y(\pi/2)}{3} & \qw \\
\lstick{\ket{01}} & \qw & \qw                    & \gate{R_y(\theta_1)}     & \qw                     & \gate{R_y(\pi+\theta_2)}     & \qw                   & \qw \\
\lstick{\ket{10}} & \qw & \qw                    & \qw                      & \gate{R_y(-4\pi/3)}     & \qw                          & \qw                   & \qw \\
\lstick{\ket{11}} & \qw & \gate{R_y(-\pi/2)}     & \qw                      & \qw                     & \qw                          & \gate{R_y(\pi/2)}     & \qw \\
} }
\caption[Two virtual qubit Hadamard ($H^{\otimes 2}$) gate.]{\textbf{Two virtual qubit Hadamard ($H^{\otimes 2}$) gate.} All transitions involve the $\ket{00}$ encoded state, as necessitated by the star-type topology of the state set chosen in Fig.~\hyperref[fig:virtqubit_BVA_results]{\ref*{fig:virtqubit_BVA_results}b}. $\theta_i = 2\arcsin(\sqrt{i/3})$.}
\label{suppfig:two_virt_hadamard}
\end{figure}

\begin{figure}
  \centering
  \mbox{ \Qcircuit @C=1em @R=2em {
\lstick{\ket{000}} & \qw & \sgate{R_y(\theta_1)}{7} & \sgate{R_y(\theta_2)}{6} & \sgate{R_y(\theta_3)}{5} & \sgate{R_y(\theta_4)}{4} & \sgate{R_y(\theta_5)}{3} & \sgate{R_y(\theta_6)}{2} & \sgate{R_y(\theta_7)}{1} & \qw \\
\lstick{\ket{001}} & \qw & \qw                      & \qw                      & \qw                      & \qw                      & \qw                      & \qw                      & \gate{R_y(\theta_7)}     & \qw \\
\lstick{\ket{010}} & \qw & \qw                      & \qw                      & \qw                      & \qw                      & \qw                      & \gate{R_y(\theta_6)}     & \qw                      & \qw \\
\lstick{\ket{011}} & \qw & \qw                      & \qw                      & \qw                      & \qw                      & \gate{R_y(\theta_5)}     & \qw                      & \qw                      & \qw \\
\lstick{\ket{100}} & \qw & \qw                      & \qw                      & \qw                      & \gate{R_y(\theta_4)}     & \qw                      & \qw                      & \qw                      & \qw \\
\lstick{\ket{101}} & \qw & \qw                      & \qw                      & \gate{R_y(\theta_3)}     & \qw                      & \qw                      & \qw                      & \qw                      & \qw \\
\lstick{\ket{110}} & \qw & \qw                      & \gate{R_y(\theta_2)}     & \qw                      & \qw                      & \qw                      & \qw                      & \qw                      & \qw \\
\lstick{\ket{111}} & \qw & \gate{R_y(\theta_1)}     & \qw                      & \qw                      & \qw                      & \qw                      & \qw                      & \qw                      & \qw \\
} }
\caption[Fast superposition pulse using 7 Givens rotations.]{\textbf{Fast superposition pulse using 7 Givens rotations.} All transitions involve the $\ket{000}$ encoded state, as necessitated by the star-type topology of the state set chosen in Fig.~\hyperref[fig:virtqubit_BVA_results]{\ref*{fig:virtqubit_BVA_results}b}. The total gate time for these 7 transitions is \SI{170}{\micro\second}. Here, $\theta_i = 2\arcsin(\sqrt{1/(d-j+1)})$.}
\label{suppfig:fast_superposition_pulse}
\end{figure}

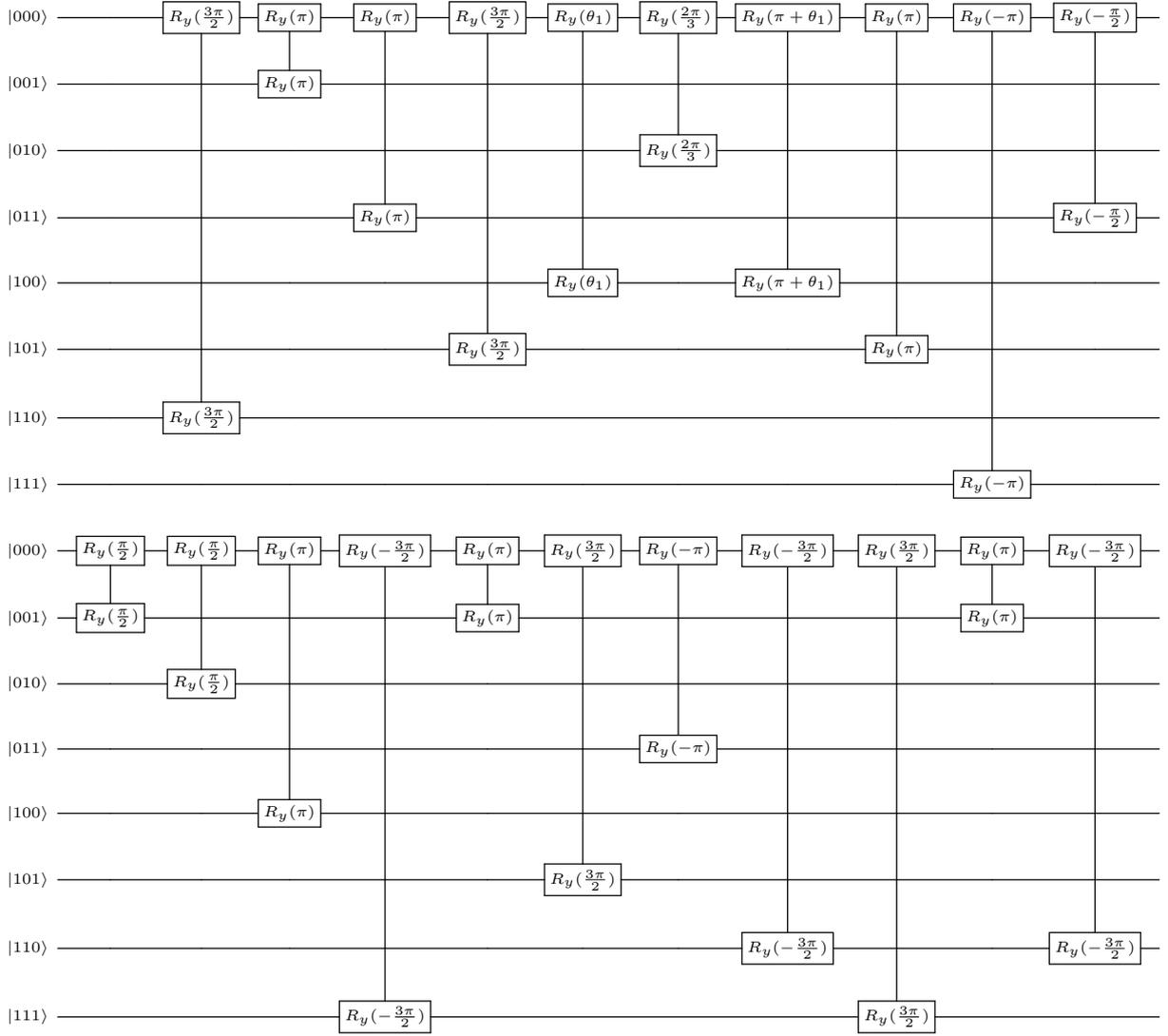
\begin{figure}
  \centering \tiny \mbox{ \Qcircuit @C=1em @R=2em {
      \lstick{\ket{000}} & \qw & \sgate{R_y(\frac{3\pi}{2})}{6} & \sgate{R_y(\pi)}{1} & \sgate{R_y(\pi)}{3} & \sgate{R_y(\frac{3\pi}{2})}{5} & \sgate{R_y(\theta_1)}{4} & \sgate{R_y(\frac{2\pi}{3})}{2} & \sgate{R_y(\pi+\theta_1)}{4} & \sgate{R_y(\pi)}{5}    & \sgate{R_y(-\pi)}{7} & \sgate{R_y(-\frac{\pi}{2})}{3} & \qw \\
      \lstick{\ket{001}} & \qw & \qw                             & \gate{R_y(\pi)}     & \qw                    & \qw                        & \qw                    & \qw                              & \qw                          & \qw                    & \qw                  & \qw                            & \qw \\
      \lstick{\ket{010}} & \qw & \qw                             & \qw                    & \qw                    & \qw                     & \qw                    &  \gate{R_y(\frac{2\pi}{3})}      & \qw                          & \qw                    & \qw                  & \qw                            & \qw \\
      \lstick{\ket{011}} & \qw & \qw                             & \qw                    &  \gate{R_y(\pi)}    & \qw                        & \qw                    & \qw                              & \qw                          & \qw                    & \qw                  & \gate{R_y(-\frac{\pi}{2})}     & \qw \\
      \lstick{\ket{100}} & \qw & \qw                             & \qw                    & \qw                    & \qw                     &  \gate{R_y(\theta_1)}    & \qw                            &  \gate{R_y(\pi+\theta_1)}    & \qw                    & \qw                  & \qw                            & \qw \\
      \lstick{\ket{101}} & \qw & \qw                             & \qw                    & \qw              & \gate{R_y(\frac{3\pi}{2})}    & \qw                    & \qw                              & \qw                          & \gate{R_y(\pi)}        & \qw                  & \qw                            & \qw \\
      \lstick{\ket{110}} & \qw & \gate{R_y(\frac{3\pi}{2})}     & \qw                    & \qw                    & \qw                      & \qw                    & \qw                              & \qw                          &  \qw                   & \qw                  & \qw                            & \qw \\
      \lstick{\ket{111}} & \qw & \qw                             & \qw                    & \qw                    & \qw                     & \qw                    & \qw                              & \qw                          & \qw                    &  \gate{R_y(-\pi)}    & \qw                            & \qw \\
      \lstick{\ket{000}} & \sgate{R_y(\frac{\pi}{2})}{1} & \sgate{R_y(\frac{\pi}{2})}{2} & \sgate{R_y(\pi)}{4} & \sgate{R_y(-\frac{3\pi}{2})}{7} & \sgate{R_y(\pi)}{1}   & \sgate{R_y(\frac{3\pi}{2})}{5} & \sgate{R_y(-\pi)}{3} & \sgate{R_y(-\frac{3\pi}{2})}{6} & \sgate{R_y(\frac{3\pi}{2})}{7} & \sgate{R_y(\pi)}{1} & \sgate{R_y(-\frac{3\pi}{2})}{6} & \qw \\
      \lstick{\ket{001}} & \gate{R_y(\frac{\pi}{2})}     & \qw                           & \qw                   & \qw                           & \gate{R_y(\pi)}       & \qw                            & \qw                   & \qw                            & \qw                            & \gate{R_y(\pi)}     & \qw                             & \qw \\
      \lstick{\ket{010}} & \qw                           &  \gate{R_y(\frac{\pi}{2})}    & \qw                   & \qw                           & \qw                   & \qw                            & \qw                   & \qw                          & \qw                            & \qw                 & \qw                               & \qw \\
      \lstick{\ket{011}} & \qw                           & \qw                           & \qw                   & \qw                           & \qw                   & \qw                            & \gate{R_y(-\pi)}     & \qw                           & \qw                            & \qw                 & \qw                               & \qw \\
      \lstick{\ket{100}} & \qw                           & \qw                           &  \gate{R_y(\pi)}    & \qw                             & \qw                   & \qw                            & \qw                   & \qw                          & \qw                            & \qw                 & \qw                               & \qw \\
      \lstick{\ket{101}} & \qw                           & \qw                           & \qw                   & \qw                           & \qw                   &  \gate{R_y(\frac{3\pi}{2})}    & \qw                   & \qw                          & \qw                            & \qw                 & \qw                               & \qw \\
      \lstick{\ket{110}} & \qw                           & \qw                           & \qw                   & \qw                           & \qw                   & \qw                            & \qw                   &  \gate{R_y(-\frac{3\pi}{2})} & \qw                            & \qw                 &  \gate{R_y(-\frac{3\pi}{2})}      & \qw \\
      \lstick{\ket{111}} & \qw                           & \qw                           & \qw                   & \gate{R_y(-\frac{3\pi}{2})}   & \qw                   & \qw                           & \qw                   & \qw                           &  \gate{R_y(\frac{3\pi}{2})}    & \qw                 & \qw                               & \qw \\
    } }
  \caption[Three virtual qubit Hadamard ($H^{\otimes 2}$) gate.]{\textbf{Three virtual qubit Hadamard ($H^{\otimes 2}$) gate.} All transitions involve the $\ket{000}$ encoded state, as necessitated by the star-type topology of the state set chosen in Fig.~\hyperref[fig:virtqubit_BVA_results]{\ref*{fig:virtqubit_BVA_results}b}. $\theta_1 =
    2\arcsin(\sqrt{1/3})$.}
  \label{suppfig:three_virt_hadamard}
\end{figure}

\section{Pulse sequences and encodings for gates and algorithms}
\label{sec:6_pulse_sequences}

\begin{figure} \centering
\includegraphics[width=0.45\textwidth]{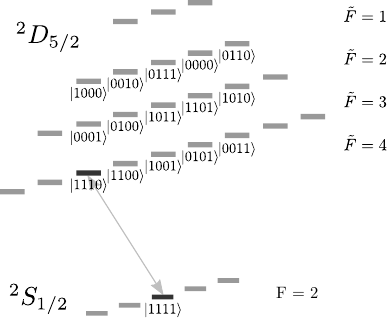}
\caption[State encoding scheme for \textit{CCCNOT} gate.]{\textbf{State encoding scheme for \textit{CCCNOT} gate.} States chosen to encode 4 virtual qubits into the ion, using transitions that are all connected via the $\ket{S_{1/2}, F=2, m_F=0}$ state to preserve the star-topology used for qudit Ramsey and Bernstein-Vazirani experiments. The grey double-headed arrow indicates the transition used for driving the single pulse required to implement this gate.}
  \label{suppfig:CCCNOT_state_encodings}
\end{figure}

Below, we explicitly show (Supp.~Fig.~\ref{suppfig:qudit_Ramsey_all_encoding_schemes}) the encoded states for every qudit Ramsey measurement from $d=2$ to 24, with colour-coded state labels that indicate which transition (i.e. from which $S_{1/2}$ state the transition is driven) is used. In Supp. figs. \ref{suppfig:two_virt_hadamard} and \ref{suppfig:three_virt_hadamard}, we show the full pulse sequences used for the implementation of the $H^{\otimes n}$ gates in the Bernstein-Vazirani algorithm. Supp.~Fig.~\ref{suppfig:fast_superposition_pulse} shows the fast \enquote{superposition pulse} used to quickly initialise the 3 virtual qubit state into $\frac{1}{\sqrt{2^n}} \sum\limits_{x=0}^{2^n-1}\ket{x}$.
Finally, in Supp.~Fig.~\ref{suppfig:CCCNOT_state_encodings} we show the states chosen to encode the basis states for 4 virtual qubits as used in the truth table measurement for the \textit{CCCNOT} gate.

\bibliographystyle{apsrev4-1}
\bibliography{references}